\documentclass[aps,prb,showpacs,twocolumn,floats]{revtex4}
\usepackage{amssymb}

\usepackage{graphicx,psfig}
\usepackage{dcolumn}
\usepackage{bm}

\input{tcilatex}

\begin{document}

\title{Towards Microscopic Understanding of the Phonon Bottleneck }
\author{D. A. Garanin}
\affiliation{\mbox{Department of Physics and Astronomy, Lehman
College, City University of New York,} \\ \mbox{250 Bedford Park
Boulevard West, Bronx, New York 10468-1589, U.S.A.} }
\date{\today}

\begin{abstract}
The problem of the phonon bottleneck in the relaxation of two-level systems
(spins) to a narrow group of resonant phonons via emission-absorption
processes is investigated from the first principles. It is shown that the
kinetic approach based on the Pauli master equation is invalid because of
the narrow distribution of the phonons exchanging their energy with the
spins. This results in a long-memory effect that can be best taken into
account by introducing an additional dynamical variable corresponding to the
nondiagonal matrix elements responsible for spin-phonon correlation. The
resulting system of dynamical equations describes the phonon-bottleneck
plateau in the spin excitation, as well as a gap in the spin-phonon spectrum
for any finite concentration of spins. On the other hand, it does not
accurately render the lineshape of emitted phonons and still needs improving.
\end{abstract}
 \maketitle


\section{Introduction}

The problem of the phonon bottleneck (PB) was recognized in 1941 by Van
Vleck who started the abstract of his article\cite{vle41pr} with the
sentence ``The present paper is rather negative in character''. Analyzing
different ways for the phonons emitted by magnetic impurities (henceforth
spins) to relax, Van Vleck found that in many typical cases the phonon rates
are insufficient to keep the phonon subsystem at equilibrium. The phonons
emitted by the \emph{direct processes}, forming a narrow resonant group in
the energy space, should be absorbed by spins again, that drastically
throttles the net relaxation rate of the latter. This happens, of course, if
the concentration of spins is large enough. There is no PB for a single spin
interacting with phonons in a macroscopic crystal, no matter what is the
phonon relaxation rate.

Van Vleck's ``negative finding'' excited other researchers for an activity
that has not ebbed until now. Subsequent publications\cite
{and59pr,gionas65pr} explored the analogy with trapping of resonant
radiation in gases, considered earlier by Holstein. \cite{hol4751pr} The
corresponding theory deals with the spatial diffusion of emitted and
reabsorbed photons or phonons from the body of the specimen towards its
boundaries where they escape (see also the recent Ref.\ \onlinecite{tam97prb}%
). Giordmaine and Nash\cite{gionas65pr} cite a number of early experiments
where the PB was observed. Recently indications of phonon bottleneck were
seen in molecular magnets.\cite{chietal00V15,chietal03prbrc} The usual
``fingerprint'' of the PB is the decrease of the net spin relaxation rate if
the thermal contact between the crystal and the holder is bad. Then the
emitted phonons cannot efficiently escape from the crystal, and the only way
of their relaxation are nonlinear phonon processes. The dependence of the PB
on \ the thermal contact becomes pronounced at low temperatures, where the
direct processes are dominating. At higher temperatures the \emph{Raman
processes} strongly come into play. It was shown\cite
{scojef62pr,brywag66pr,gil73jpc} that a kind of phonon bottleneck takes
place for the Raman processes as well, but the effect is much weaker than
that for the direct processes.

Existing theories of the phonon bottleneck for the direct processes operate
with diffusion, kinetic, or rate equations that are not derived from the
first principles. These equations are set up using the balance of the
excitation numbers and the energy between the spins and phonons, as well as
the Fermi golden rule. The quasicontinuum of resonant phonons is considered
in a simplified way as a single dynamical variable. One group of
publications \cite{hol4751pr,gionas65pr,tam97prb} takes into account effects
of spatial inhomogeneity in the sample. Another group \cite
{faustr61jpcs,scojef62pr,brywag67pr,abrble70,pinfai90prb} further simplifies
the problem by ignoring spatial effects and modelling the relaxation of
phonons by a single rate of approaching the equilibrium at some bath
temperature.

Related class of problems deals with the influence of spins on the phonon
spectrum\cite{jacste63pr} and with the resonance scattering of acoustic
waves on spins (see Ref.\ \onlinecite{lai86prb} and references therein).
Jacobsen and Stevens\cite{jacste63pr} shown that the hybridization between
spins and phonons in a crystal with \emph{regularly spaced} spins leads to a
\emph{gap} or \emph{stop-band} in the phonon spectrum around the spin-phonon
resonance. This can be considered as an extreme case of the phonon
bottleneck, as, obviously, one cannot speak about a unidirectional energy
transfer from spins to phonons, at least if the phonons are undamped. If
spins are diluted, the resonance phonon gap should persist, in a reduced
form. This effect was never discussed in the existing theories of the PB.

Since Van Vleck's originating work, the main stress in the PB problem was
put on the sufficiency or insufficiency of the phonon relaxation processes
to transport the energy away from the spins. A common idea is that the
phonon bottleneck is something that happens if the phonons are not relaxing
fast enough. One can find in the literature a dimensionless \emph{bottleneck
factor} that becomes large if the phonon relaxation rate is small [see,
e.g., Eq.\ (36) of Ref.\ \onlinecite{scojef62pr}]. However, numerous
publications on the \emph{single-spin-plus-phonon-bath} model do not care
for the phonon relaxation at all, that is certainly correct. The problem
arises only if the number of spins is macroscopic and their concentration is
large enough. Then the PB occurs and, as a consequence, one has to take into
account the relaxation of phonons. This means that there should be another
dimensionless \emph{bottleneck parameter} that is independent of the phonon
relaxation or escape rate.

The important component of the PB problem dealing exclusively with the
energy transfer between the spins and resonant phonons has not recieved a
due attention until now. Starting, for simplicity, with undamped harmonic
phonons, one can ask what will be the accurate dynamics of the system, the
large-time asymptote of the evolution, the energy distribution of emitted
phonons under the bottleneck condition. One can expect that the system of
equations\cite{faustr61jpcs,scojef62pr,brywag67pr,abrble70,pinfai90prb}
describing resonant phonons as a single dynamical variable follows from a
more \emph{detailed} energy-resolved theoretical framework via the
integration over the phonon energies.

Such a detailed description is likely to include an equation of motion for
the spin averages coupled to the system of kinetic equations for the phonons
of all possible modes, considered separately. Are there any difficulties
that prevent the derivation of such detailed spin-phonon equations from the
first principles? It is puzzling why it has not been done yet, given that
quantum kinetic equations result from the many-body quantum mechanics, under
certain conditions. The procedure includes at first the derivation of the
Pauli master equation\cite{pau28,shepri60pr} that breaks the reversibility
of the quantum mechanics. Then kinetic equations for particular observables
can be obtained by averaging appropriate operators with the Pauli master
equation.

Studying the mechanism of the energy exchange between the spins and the
resonant group of phonons is the aim of this work. Identical spins $S=1/2$
(two-level systems) randomly placed in the crystal are considered. One
starts with undamped harmonic phonons and adds their relaxation at the very
end in the same simplest way as was done in Refs.\
\onlinecite
{faustr61jpcs,scojef62pr,brywag67pr,abrble70,pinfai90prb}. The main finding
is that the narrow distribution of emitted phonons violates the condition
for the Pauli master equation to be valid. Hence no standard kinetic
equations for the resonant phonons can be derived. Instead, one obtains on
this way equations with memory that are of little practical use. In this
situation it is better to step back and use a more basic system of dynamic
equations including nondiagonal elements of the spin-phonon density matrix.
Numerical solution of this system of equations shows a phonon bottleneck for
the concentration of spins large enough. There is a nontrivial asymptotic
plateau for the spin excitation, if the phonons are undamped. For a low
level of the initial spin excitation the resulting system of equations can
be solved analytically. The analytical solution shows a resonance gap that
is similar to that found by Jacobsen and Stevens\cite{jacste63pr} for
nondiluted spins. For the model with the damped phonons, the system
eventually relaxes to the thermal equilibrium. However, the effective
relaxation rate for the spins is much less than the phonon relaxation rate
in the case of a strong bottleneck, contrary to the first-glance expectation.

The following part of the article is organized as follows. Sec.\ \ref
{Sec-Hamiltonian} sets up the Hamiltonian of the spin-phonon system, its the
quantum states and the Schr\"{o}dinger equation. In Sec.\ \ref{Sec-MasterEq}
the derivation of the Pauli master equation is reviewed and its
applicability conditions are discussed. It is argued that for a narrow
resonant group of phonons the standard kinetic formalism based on the master
equation should fail because of the long-memory effect. Sec.\ \ref
{Sec-OriginBott} contains the discussion of the conditions for the phonon
bottleneck, and the dimensionless \emph{bottleneck parameter} is introduced.
Sec.\ \ref{Sec-Short-memory} explores the short-memory approach to the PB
based on the master equation. One obtains a system of coupled kinetic
equations for spins and phonons that contain powers of the energy $\delta $%
-function and is thus incorrect. It is shown how one can obtain previously
published bottleneck equations\cite
{faustr61jpcs,scojef62pr,brywag67pr,abrble70,pinfai90prb} by mathematically
non-rigorous manipulations with $\delta $-functions. In Sec.\ \ref
{Sec-PB-Dynamical} one steps back to the master equation with memory and
obtains a system of \emph{dynamical} eqiations describing the
emission/absorption of phonons by spins and the bottleneck. These equations
are enchanced by including terms responsible for the relaxation of phonons
in a simple way. Numerical solutions are presented and discussed. Further
the analytical solution of the system of dynamical spin-phonon equations in
the case of low spin excitation is presented, and analytical results for the
bottleneck plateau in the spin axcitation and the effective relaxation rate
of the spin-phonon system under the bottleneck condition are obtained.
Discussion is done throughout the article. Sec.\ \ref{Sec-Summary} contains
a summary of the results obtained, as well as a discussion of the
inhomogeneous broadening and the interplay between the phonon bottleneck and
phonon superradiance. \cite{chugar04prl}

\section{The Hamiltonian and Schr\"{o}dinger equation}
\label{Sec-Hamiltonian}

Consider a spin-phonon Hamiltonian for $N_{S}$ two-level systems (spins)
within an elastic body of $N$ cells
\begin{equation}
\hat{H}=\hat{H}_{0}+\hat{V},  \label{HamFull}
\end{equation}
where
\begin{equation}
\hat{H}_{0}=-\frac{\hbar \Bbb{\omega }_{0}}{2}\sum_{i}\sigma _{iz}+\sum_{%
\mathbf{k}}\hbar \omega _{\mathbf{k}}a_{\mathbf{k}}^{\dagger }a_{\mathbf{k}}
\label{H0Def}
\end{equation}
describes spins and harmonic phonons, $\mathbf{\sigma }$ being the Pauli
matrix. Neglecting the processes that do not conserve the energy, one can
write this Hamiltonian in the rotating-wave approximation as
\begin{equation}
\hat{V}=-\frac{\hbar }{\sqrt{N}}\sum_{i}\sum_{\mathbf{k}}\left( A_{i\mathbf{k%
}}^{\ast }X_{i}^{01}a_{\mathbf{k}}^{\dagger }+A_{i\mathbf{k}}X_{i}^{10}a_{%
\mathbf{k}}\right) ,  \label{VRWA}
\end{equation}
where $A_{i\mathbf{k}}\equiv V_{\mathbf{k}}e^{-i\mathbf{k\cdot r}_{i}}.$
Below $\hat{V}$ will be treated as a perturbation. The operator $%
X^{10}\equiv \sigma _{-}$ brings the spin from the ground state $\left|
\uparrow \right\rangle \equiv \left| 0\right\rangle \equiv \binom{1}{0}$ to
the excited state $\left| \downarrow \right\rangle \equiv \left|
1\right\rangle \equiv \binom{0}{1}$ while $X^{01}\equiv \sigma _{+}$ does
the opposite.

To describe quantum states of the system of phonons and spins, we use a
basis that is a direct product of the states of all phonon modes and of the
spins. The wave function can be written in the form
\begin{equation}
\Psi =\sum_{\left\{ \mu _{i}\right\} \left\{ \nu _{\mathbf{k}}\right\}
}c_{\left\{ \mu _{i}\right\} \left\{ \nu _{\mathbf{k}}\right\}
}\prod_{i}\left| \mu _{i}\right\rangle \prod_{\mathbf{k}}\left| \nu _{%
\mathbf{k}}\right\rangle ,  \label{PsiGeneral}
\end{equation}
where $\mu _{i}=0,1$ corresponding to the ground and excited states of the
spins, respectively, and $\nu _{\mathbf{k}}=0,1,2,\ldots $ are the
occupation numbers of the phonon modes. We will use the shortcuts for
different basis states
\begin{equation}
\left\{ \mu _{i}\right\} \equiv \mathcal{S},\quad \left\{ \nu _{\mathbf{k}%
}\right\} \equiv \mathcal{P},\quad \left\{ \mu _{i}\right\} \left\{ \nu _{%
\mathbf{k}}\right\} =\left( \mathcal{SP}\right) \equiv \mathcal{W.}
\label{Shortcuts}
\end{equation}
The Schr\"{o}dinger equation (SE) for our spin-phonon system $i\hbar d\Psi
/dt=\hat{H}\Psi $ can be written as a system of equations for the
coefficients $c_{\mathcal{W}}$ that can be obtained by acting on $\Psi $
with the Hamiltonian of Eq.\ (\ref{HamFull}). The general form of the SE for
the coefficients is
\begin{eqnarray}
i\hbar \frac{d}{dt}c_{\mathcal{W}} &=&\sum_{\mathcal{W}_{1}}c_{\mathcal{W}%
_{1}}\left\langle \mathcal{W}\left| \hat{H}\right| \mathcal{W}%
_{1}\right\rangle  \nonumber \\
&=&\varepsilon _{\mathcal{W}}c_{\mathcal{W}}+\sum_{\mathcal{W}_{1}}c_{%
\mathcal{W}_1}\left\langle \mathcal{W}\left| \hat{V}\right| \mathcal{W}%
_{1}\right\rangle ,  \label{SEtcGen}
\end{eqnarray}
where
\begin{equation}
\varepsilon _{\mathcal{W}}=\left\langle \mathcal{W}\left| \hat{H}_{0}\right|
\mathcal{W}\right\rangle =\hbar \omega _{\mathcal{W}}  \label{epsWDef}
\end{equation}
is the unperturbed energy in the state $\mathcal{W}$. It is convenient to
introduce the slow amplitudes $\tilde{c}_{\mathcal{W}}$ via
\begin{equation}
c_{\mathcal{W}}(t)=\tilde{c}_{\mathcal{W}}(t)e^{-i\omega _{\mathcal{W}}t},
\end{equation}
then the SE becomes
\begin{equation}
\frac{d}{dt}\tilde{c}_{\mathcal{W}}=-\frac{i}{\hbar }\sum_{\mathcal{W}%
_{1}}e^{i\left( \omega _{\mathcal{W}}-\omega _{\mathcal{W}_{1}}\right) t}%
\tilde{c}_{\mathcal{W}_{1}}\left\langle \mathcal{W}\left| \hat{V}\right|
\mathcal{W}_{1}\right\rangle .  \label{SESlow}
\end{equation}

Let us work out the concrete form of the SE, for the spin-phonon interaction
$\hat{V}$ given by Eq.\ (\ref{VRWA}). In the matrix element $\left\langle
\mathcal{W}\left| \hat{V}\right| \mathcal{W}_{1}\right\rangle $ the state $%
\mathcal{W}_{1}$ differs from $\mathcal{W}$ by one spin flip and creation or
annihilation of one phonon. Thus it is convenient to write $\mathcal{W}_{1}$
in the incremental form$.$ In particular, $\mathcal{W}_{1}=\mathcal{W}%
,1_{i},-1_{\mathbf{k}}$ means that in $\mathcal{W}_{1}$ the spin on the site
$i$ is excited and one phonon in the $\mathbf{k}$-mode is annihilated,
relative to $\mathcal{W.}$ This can only happen if $\mu _{i}=0$ in $\mathcal{%
W.}$ In the state $\mathcal{W}_{1}=\mathcal{W},0_{i},+1_{\mathbf{k}} $ the
spin on the site $i$ is deexcited and one phonon in the $\mathbf{k}$-mode is
created, relative to $\mathcal{W.}$ This can only happen if $\mu _{i}=1$ in $%
\mathcal{W.}$ One obtains
\begin{eqnarray}
\left\langle \mathcal{W}\left| \hat{V}\right| \mathcal{W},0_{i},+1_{\mathbf{k%
}}\right\rangle &=&-\frac{\hbar }{\sqrt{N}}A_{i\mathbf{k}}\mu _{i}\sqrt{\nu
_{\mathbf{k}}+1}  \nonumber \\
\left\langle \mathcal{W}\left| \hat{V}\right| \mathcal{W},1_{i},-1_{\mathbf{k%
}}\right\rangle &=&-\frac{\hbar }{\sqrt{N}}A_{i\mathbf{k}}^{\ast }\left(
1-\mu _{i}\right) \sqrt{\nu _{\mathbf{k}}},  \label{VME}
\end{eqnarray}
where $\mu _{i}$ and $\nu _{\mathbf{k}}$ refer to the state $\mathcal{W.}$
With the help of this, Eq.\ (\ref{SESlow}) can be written in the form
\begin{eqnarray}
&&\frac{d\tilde{c}_{\mathcal{W}}}{dt}=\frac{i}{\sqrt{N}}\sum_{i}\sum_{%
\mathbf{k}}e^{-i\left( \omega _{\mathbf{k}}-\omega _{0}\right) t}A_{i\mathbf{%
k}}\mu _{i}\sqrt{\nu _{\mathbf{k}}+1}\tilde{c}_{\mathcal{W},0_{i},+1_{%
\mathbf{k}}}  \nonumber \\
&&{}+\frac{i}{\sqrt{N}}\sum_{i}\sum_{\mathbf{k}}e^{i\left( \omega _{\mathbf{k%
}}-\omega _{0}\right) t}A_{i\mathbf{k}}^{\ast }\left( 1-\mu _{i}\right)
\sqrt{\nu _{\mathbf{k}}}\tilde{c}_{\mathcal{W},1_{i},-1_{\mathbf{k}}}.
\label{SEkmu}
\end{eqnarray}

\section{The Pauli master equation}
\label{Sec-MasterEq}

Let us construct now the elements of the slow density-matrix for our closed
system of spins and phonons by multiplying coefficients of the wave function
\begin{equation}
\tilde{\rho}_{\mathcal{W|W}^{\prime }}=\tilde{c}_{\mathcal{W}}\tilde{c}_{%
\mathcal{W}^{\prime }}^{\ast }.  \label{rhoSSprDef}
\end{equation}
The equation of motion for them follows from Eq.\ (\ref{SESlow}) and reads
\begin{eqnarray}
&&\frac{d}{dt}\tilde{\rho}_{\mathcal{W|W}^{\prime }}=-\frac{i}{\hbar }\sum_{%
\mathcal{W}_{1}}\left\langle \mathcal{W}\left| \hat{V}\right| \mathcal{W}%
_{1}\right\rangle e^{i\left( \omega _{\mathcal{W}}-\omega _{\mathcal{W}%
_{1}}\right) t}\tilde{\rho}_{\mathcal{W}_{1}\mathcal{|W}^{\prime }}
\nonumber \\
&&{}+\frac{i}{\hbar }\sum_{\mathcal{W}_{1}^{\prime }}\left\langle \mathcal{W}%
^{\prime }\left| \hat{V}\right| \mathcal{W}_{1}^{\prime }\right\rangle
^{\ast }e^{-i\left( \omega _{\mathcal{W}^{\prime }}-\omega _{\mathcal{W}%
_{1}^{\prime }}\right) t}\tilde{\rho}_{\mathcal{W|W}_{1}^{\prime }}.
\label{DMGenEqc}
\end{eqnarray}

We restrict our consideration to the case of a nearly diagonal density
matrix of the spin-phonon system. This will be the case if the spins are
prepared in the initial state with random phases, so that the averages of
the transverse spin components are zero, $\left\langle S_{x}\right\rangle
=\left\langle S_{y}\right\rangle =0,$ and the phonons are at thermal
equilibrium. We are going to derive equations for the populations of the
quantum states
\begin{equation}
P_{\mathcal{W}}=\tilde{\rho}_{\mathcal{W|W}}.  \label{PPopDef}
\end{equation}
One can see from Eqs.\ (\ref{DMGenEqc}) with $\mathcal{W}^{\prime }=\mathcal{%
W}$ that diagonal elements $\tilde{\rho}_{\mathcal{W|W}}$ are dynamically
coupled to nondiagonal elements such as $\tilde{\rho}_{\mathcal{W}_{1}%
\mathcal{|W}}$. The latter are generated by $\hat{V}$ and thus are small.
One can integrate them out using the equations similar to Eq.\ (\ref
{DMGenEqc}). This yields
\begin{eqnarray}
&&\tilde{\rho}_{\mathcal{W}_{1}\mathcal{|W}}(t)=\tilde{\rho}_{\mathcal{W}_{1}%
\mathcal{|W}}(t_{0})  \nonumber \\
&&-\frac{i}{\hbar }\sum_{\mathcal{W}_{2}}\left\langle \mathcal{W}_{1}\left|
\hat{V}\right| \mathcal{W}_{2}\right\rangle \int_{t_{0}}^{t}dt^{\prime
}e^{i\left( \omega _{\mathcal{W}_{1}}-\omega _{\mathcal{W}_{2}}\right)
t^{\prime }}\tilde{\rho}_{\mathcal{W}_{2}\mathcal{|W}}(t^{\prime })
\nonumber \\
&&+\frac{i}{\hbar }\sum_{\mathcal{W}_{2}}\left\langle \mathcal{W}\left| \hat{%
V}\right| \mathcal{W}_{2}\right\rangle ^{\ast }e^{-i\left( \omega _{\mathcal{%
W}}-\omega _{\mathcal{W}_{2}}\right) t}\tilde{\rho}_{\mathcal{W}_{1}\mathcal{%
|W}_{2}}
\end{eqnarray}
and $\tilde{\rho}_{\mathcal{W|W}_{1}}(t)=\left( \tilde{\rho}_{\mathcal{W}_{1}%
\mathcal{|W}}(t)\right) ^{\ast }.$ Remaining within the second order in the
perturbation $\hat{V},$ it is sufficient to drop the nondiagonal terms in
the right-hand sides of the above equation. Plugging the resulting
expressions for $\tilde{\rho}_{\mathcal{W}_{1}\mathcal{|W}}(t)$ and $\tilde{%
\rho}_{\mathcal{W|W}_{1}}(t)$ into Eq.\ (\ref{DMGenEqc}) yields the master
equation with memory
\begin{eqnarray}
&&\frac{d}{dt}P_{\mathcal{W}}(t)=\frac{2}{\hbar ^{2}}\sum_{\mathcal{W}%
_{1}}\left| \left\langle \mathcal{W}\left| \hat{V}\right| \mathcal{W}%
_{1}\right\rangle \right| ^{2}\int_{t_{0}}^{t}dt^{\prime }  \nonumber \\
&&\times \cos \left[ \left( \omega _{\mathcal{W}}-\omega _{\mathcal{W}%
_{1}}\right) \left( t-t^{\prime }\right) \right] \left[ P_{\mathcal{W}%
_{1}}(t^{\prime })-P_{\mathcal{W}}(t^{\prime })\right] .  \label{NEqMemory}
\end{eqnarray}

In the usual case where the kinetic theory is applicable, $\left|
\left\langle \mathcal{W}\left| \hat{V}\right| \mathcal{W}_{1}\right\rangle
\right| ^{2}$ and $P_{\mathcal{W}_{1}}(t^{\prime })$ are smooth functions of
$\mathcal{W}_{1}.$ The summation over $\mathcal{W}_{1}$ in Eq.\ (\ref
{NEqMemory}) goes over a wide energy interval limited by the maximal energy
of the phonon reservoire $\hbar \omega _{\max }$ in some terms and limited
by the thermal energy $k_{B}T$ in other terms. The resulting expression is
then peaked at $t-t^{\prime }\lesssim 1/\omega _{\max }$ or $t-t^{\prime }
\lesssim \hbar /(k_{B}T).$ Both of these characteristic times are much
shorter than the relaxation time $1/\Gamma $ for $P_{\mathcal{W}}(t)$ since
the spin-phonon relaxation rate $\Gamma \varpropto \left| V_{\mathbf{k}%
}\right| ^{2}$ is small. Thus one can make the short-memory approximation $%
P_{\mathcal{W}}(t^{\prime })\Rightarrow P_{\mathcal{W}}(t)$ and $P_{\mathcal{%
W}_{1}}(t^{\prime })\Rightarrow P_{\mathcal{W}_{1}}(t)$ in Eq.\ (\ref
{NEqMemory}). After that the time integration can be done easily,
\begin{equation}
\int_{t_{0}}^{t}dt^{\prime }\cos \left[ \left( \omega _{\mathcal{W}}-\omega
_{\mathcal{W}_{1}}\right) \left( t-t^{\prime }\right) \right] =\frac{\sin %
\left[ \left( \omega _{\mathcal{W}_{1}}-\omega _{\mathcal{W}}\right) \left(
t-t_{0}\right) \right] }{\omega _{\mathcal{W}_{1}}-\omega _{\mathcal{W}}}.
\label{CosINtegral}
\end{equation}
In the kinetic time range
\begin{equation}
t-t_{0}\sim \frac{1}{\Gamma }\gg \frac{1}{\omega _{\max }},\frac{\hbar }{%
k_{B}T}  \label{kinetict}
\end{equation}
one can replace
\begin{equation}
\frac{\sin \left[ \left( \omega _{\mathcal{W}_{1}}-\omega _{\mathcal{W}%
}\right) \left( t-t_{0}\right) \right] }{\omega _{\mathcal{W}_{1}}-\omega _{%
\mathcal{W}}}\Rightarrow \pi \delta \left( \omega _{\mathcal{W}_{1}}-\omega
_{\mathcal{W}}\right)  \label{deltafident}
\end{equation}
that breaks the time reversibility of the quantum mechanics and leads to the
famous Pauli master equation\cite{pau28}
\begin{equation}
\frac{d}{dt}P_{\mathcal{W}}=\sum_{\mathcal{W}_{1}}\tilde{\Gamma}_{\mathcal{WW%
}_{1}}\left( P_{\mathcal{W}_{1}}-P_{\mathcal{W}}\right)  \label{PWEq}
\end{equation}
with the Fermi-golden-rule detailed transition rate
\begin{equation}
\tilde{\Gamma}_{\mathcal{WW}_{1}}=\frac{2\pi }{\hbar ^{2}}\left|
\left\langle \mathcal{W}\left| \hat{V}\right| \mathcal{W}_{1}\right\rangle
\right| ^{2}\delta \left( \omega _{\mathcal{W}_{1}}-\omega _{\mathcal{W}%
}\right) .  \label{GammatildeWW1}
\end{equation}
The sum of the latter over $\mathcal{W}_{1}$
\begin{equation}
\Gamma _{\mathcal{W}}=\sum_{\mathcal{W}_{1}}\tilde{\Gamma}_{\mathcal{WW}%
_{1}}=\frac{2\pi }{\hbar ^{2}}\sum_{\mathcal{W}_{1}}\left| \left\langle
\mathcal{W}\left| \hat{V}\right| \mathcal{W}_{1}\right\rangle \right|
^{2}\delta \left( \omega _{\mathcal{W}_{1}}-\omega _{\mathcal{W}}\right)
\label{GammaFermigr}
\end{equation}
is the decay rate of the state $\mathcal{W.}$ One can write $\Gamma _{%
\mathcal{W}}$ as
\begin{equation}
\Gamma _{\mathcal{W}}=\frac{2\pi }{\hbar ^{2}}N\left\langle \left|
\left\langle \mathcal{W}\left| \hat{V}\right| \mathcal{W}_{1}\right\rangle
\right| ^{2}\right\rangle \rho \left( \omega _{\mathcal{W}}\right) ,
\label{Gammaviarho}
\end{equation}
where $N$ is the total number of phonon modes in the system that is
proportional to the number of atoms in it, $\left\langle \left| \left\langle
\mathcal{W}\left| \hat{V}\right| \mathcal{W}_{1}\right\rangle \right|
^{2}\right\rangle $ is the average over the resonant states $\mathcal{W}%
_{1}, $ and
\begin{equation}
\rho \left( \omega _{\mathcal{W}}\right) =\frac{1}{N}\sum_{\mathcal{W}%
_{1}}\delta \left( \omega _{\mathcal{W}_{1}}-\omega _{\mathcal{W}}\right)
\label{rhoomega}
\end{equation}
is the density of states that satisfies $\int d\omega _{\mathcal{W}}\rho
\left( \omega _{\mathcal{W}}\right) =1.$ The name ``master equation'' says
that one can generate kinetic equations for different physical quantities
from it by averaging appropriate operators over the quantum states $\mathcal{%
W}$ with $P_{\mathcal{W}}.$

The reader can find a more extensive derivation and analysis of the Pauli
master equation in Ref.\ \onlinecite{shepri60pr}. The authors argue that, as
Eq.\ (\ref{PWEq}) indeed suggests, a system initially in the quantum state $%
\mathcal{W}$ will spread over all mutually accessible resonant states $%
\mathcal{W}_{1}$ in a nonoscillative way, the final result of the relaxation
being the microcanonical distribution $P_{\mathcal{W}_{1}}\varpropto \delta
\left( \omega _{\mathcal{W}_{1}}-\omega _{\mathcal{W}}\right) .$ However
plausible these arguments might appear, there is a problem since one obtains
a square of the energy $\delta $-function in Eq.\ (\ref{PWEq}) in the case
of the decay of an initial fully occupied state $\mathcal{W}$: One $\delta $%
-function is contained in the transition probability $\tilde{\Gamma}_{%
\mathcal{WW}_{1}}$ of Eq.\ (\ref{GammatildeWW1}) and another one is carried
by $P_{\mathcal{W}_{1}}.$ In the derivation of the master equation above
[see discussion below Eq.\ (\ref{NEqMemory})] it was stressed that $P_{%
\mathcal{W}_{1}}(t^{\prime })$ should be a smooth functions of $\mathcal{W}%
_{1}$ for the short-memory approximation leading to Eq.\ (\ref{PWEq}) to be
valid. Physically it means that the probability of the quantum system should
be not sharp but distributed over many states with different energies. Only
in this case one can rigirously derive kinetic equations. In the case of the
spin-phonon relaxation via direct processes, the energy of the system is
fixed and the emitted phonons build a narrow resonant group in the energy
space that is as sharp as the energy $\delta $-function describing the
probability of this process. It will be demonstrated below that this results
in the inapplicability of the kinetic approach to the description of the
phonon bottleneck and that effective bottleneck equations of Refs.\
\onlinecite
{faustr61jpcs,scojef62pr,brywag67pr,abrble70,pinfai90prb} can only be
obtained by mathematically incorrect manipulations with $\delta $-functions.

\section{The origin of the phonon bottleneck}
\label{Sec-OriginBott}

As we have seen above, the Pauli master equation does not resolve lineshapes
that is the origin of the difficulties of applying it to the PB problem.
Finite linewidths, however, follow from the time-energy uncertainty
principle. In the simple case of one spin in a macroscopic crystal (or one
atom in a free space or a large cavity) one can neglect the incoming term in
Eq.\ (\ref{PWEq}), then the solution for the decay of the initially prepared
state is $P_{\mathcal{W}}(t)=e^{-\Gamma _{\mathcal{W}}t}.$ The finite
lifetime of the decaying state leads to the finite linewidth of the emitted
phonons. At $t\rightarrow \infty $ from the Schr\"{o}dinger equation one
obtains\cite{weiwig30a,hei54}
\begin{equation}
P_{\mathcal{W}_{1}}=\frac{1}{\hbar ^{2}}\left| \left\langle \mathcal{W}%
\left| \hat{V}\right| \mathcal{W}_{1}\right\rangle \right| ^{2}\frac{1}{%
\left( \omega _{\mathcal{W}_{1}}-\omega _{\mathcal{W}}\right) ^{2}+\Gamma _{%
\mathcal{W}}^{2}/4}  \label{PWSE}
\end{equation}
that satisfies $\sum_{\mathcal{W}_{1}}P_{\mathcal{W}_{1}}=1.$

\begin{figure}[t]
\unitlength1cm
\begin{picture}(11,4)
\centerline{\psfig{file=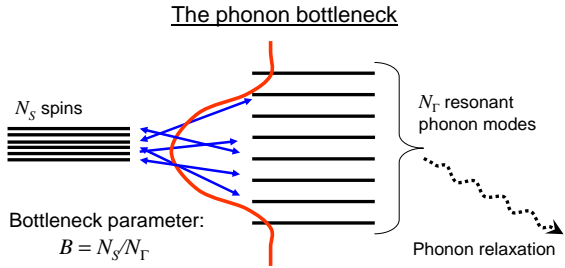,angle=-90,width=12cm}}
\end{picture}
\caption{The phonon bottleneck happens for $B\gtrsim 1$ and it results in
the reabsorption of the initially emitted phonons by spins. Fast phonon
relaxation supresses the bottleneck.}
\label{Fig-bottleneck}
\end{figure}

Apart of the Lorentzian dependence on the energy, $P_{\mathcal{W}_{1}}$ also
depends on the matrix element that can differ for different directions of
the phonon wave vector k$.$ One can average $P_{\mathcal{W}_{1}}$ keeping
the frequency $\omega _{\mathcal{W}_{1}}$ constant. Then using Eq.\ (\ref
{Gammaviarho}) one can rewrite the result in the form
\begin{equation}
\left\langle P_{\mathcal{W}_{1}}\right\rangle =\frac{2}{\pi N\rho \left(
\omega _{\mathcal{W}}\right) \Gamma _{\mathcal{W}}}\frac{\Gamma _{\mathcal{W}%
}^{2}/4}{\left( \omega _{\mathcal{W}_{1}}-\omega _{\mathcal{W}}\right)
^{2}+\Gamma _{\mathcal{W}}^{2}/4}  \label{PWvalueEstim}
\end{equation}
that allows to estimate the probabilities of the states with emitted
phonons. At resonance the Lorentzian factor here equals to one, so that the
estimation is $\left\langle P_{\mathcal{W}_{1}}\right\rangle \sim
1/N_{\Gamma },$ where
\begin{equation}
N_{\Gamma }=\pi N\rho \left( \omega _{\mathcal{W}}\right) \Gamma _{\mathcal{W%
}}  \label{NGammaDef}
\end{equation}
is the effective number of states in resonance with the state $\mathcal{W},$
taking into account its finite linewidth. $N_{\Gamma }\gg 1$ is the
condition for using quasicontinuum approximation for the states of the
system and replacing the sums over $\mathcal{W}$ by integrals. Only if $%
N_{\Gamma }\gg 1$\ is fulfilled, one can write $\delta \left( \omega _{%
\mathcal{W}_{1}}-\omega _{\mathcal{W}}\right) $ in Eqs.\ (\ref{GammatildeWW1}%
) and (\ref{rhoomega}).

For macroscopic bodies the factor $N$ in Eq.\ (\ref{NGammaDef}) makes $%
N_{\Gamma }$ very large indeed. In this case $\left\langle P_{\mathcal{W}%
}\right\rangle $ of Eq.\ (\ref{PWvalueEstim}) is very small. This validates
neglecting the incoming terms in the master equation that leads to $P_{%
\mathcal{W}}(t)=e^{-\Gamma _{\mathcal{W}}t}$ and Eq.\ (\ref{PWSE}). The
decay of an initially prepared state into the continuum can be understood
from a very simple statistical argument. The excitation initially localized
at the state $\mathcal{W}_{0}$ becomes equidistributed between the $%
N_{\Gamma }+1\cong N_{\Gamma }$ states as the result of the relaxation. Then
the probability to find the system in any of the states $1/N_{\Gamma }$ is
extremely small. Once the excited state of the spin has decayed, the
excitation never comes back because it is statistically unprobable.

For very small crystals the phonon modes can be so sparse that $N_{\Gamma }$
is of order one or even smaller. If the spin is at resonance with only few
phonon modes, the dynamics of the system is complicated and the spin cannot
relax completely since it is being excited by phonons again. This is the
case of the phonon bottleneck. The simplest case of the PB is that of the
resonance between the spin and a single phonon mode that is described in
quantum-mechanics textbooks as the resonance between two general quantum
mechanical states. It is well known that the probability to stay in the
intially prepared state is harmonically oscillating and there is no
relaxation.

The consideration above in this section pertains to the case of one
two-level system (spin) relaxing to a phonon bath at zero temperature. Below
we will consider, in particular, relaxation of $N_{S}$ spins 1/2 prepared in
their excited states and relaxing to a phonon bath at $T=0$ (see Fig.\ \ref
{Fig-bottleneck}). The initial state $\mathcal{W}$ (all spins excited, no
phonons) decays, initially, into the states with one spin flipped and one
phonon created. All arguments above are still valid for this process.
However, the resulting states will further decay into states with more spins
flipped and more phonons created. The number of emitted phonons can be by a
factor $N_{S}$ \ greater than that in the case of one spin. Then, in turn,
these phonons will begin to excite the spin subsystem. In this situation one
cannot neglect incoming terms in the master equation. A criterion for this
can be obtained by the generalization of the statistical argument above: For
$N_{S}$ spins at resonance with $N_{\Gamma }$ phonon modes one can construct
a dimensionless parameter
\begin{equation}
B\equiv \frac{N_{S}}{N_{\Gamma }}=\frac{N_{S}}{\pi N\rho \left( \omega _{%
\mathcal{W}}\right) \Gamma _{\mathcal{W}}}  \label{BDef}
\end{equation}
that controls the phonon bottleneck and can be called the \emph{bottleneck
parameter}. Note that here $n_{S}\equiv N_{S}/N$ is the number of spins per
unit cell, so that $B$ is independent of the system size. If $B\lesssim 1,$
then the most energy goes to phonons. If $B\gtrsim 1,$ then the energy
mainly remains in the spin subsystem, if the resonant phonons do not
transfer their energy elsewhere. The latter case corresponds to the PB. It
is clear that in most situations the inequality $B\gg 1$ should be fulfilled$%
.$ Indeed, if one roughly replaces $\rho \left( \omega _{\mathcal{W}}\right)
\Rightarrow 1/\omega _{\max },$ then $B\gg 1$ is equivalent to $\Gamma _{%
\mathcal{W}}/\omega _{\max }\ll n_{S}.$ The left part of this inequality is
very small, so that the inequality holds for any not too low concentrations
of spins.

It is tempting to take into account that the state resulting from the decay
of the spin, $\mathcal{W}_{1},$ has its own relaxation rate $\Gamma _{%
\mathcal{W}_{1}}$ due to the damping of the emitted phonons. This would lead
to the replacement $\Gamma _{\mathcal{W}}\Rightarrow \Gamma _{\mathcal{W}%
}+\Gamma _{\mathcal{W}_{1}}$ in Eq. (\ref{BDef}), where $\Gamma _{\mathcal{W}%
}=\Gamma $, the spin relaxation rate due to the phonon emission, and $\Gamma
_{\mathcal{W}_{1}}=\Gamma _{\mathrm{ph}}$, the phonon relaxation rate. Then
one can assume that the bottleneck should disappear in the case $\Gamma _{%
\mathrm{ph}}\gg \Gamma .$ Below the reader will see that the sum $\Gamma
+\Gamma _{\mathrm{ph}}$ does not arise in the theory. The PB, indeed, is
suppressed for large $\Gamma _{\mathrm{ph}},$ but the condition for this is
more subtle and given by Eq. (\ref{GammaphStrong}).

Inhomogeneous broadening that typically exceeds relaxation rates tends to
suppress the PB. Although it would be important for experiments, appropriate
extension of the theory will not be explicitly made here, for the sake of
transparency. This extension is discussed in the concluding section of the
paper.

\section{Short-memory approach to the phonon bottleneck}
\label{Sec-Short-memory}

In this section we try to derive equations describing the phonon bottleneck
from the Pauli master equation Eq.\ (\ref{PWEq}), i.e., within the
short-memory approximation. We will see that on this way one has to deal
with powers of the energy $\delta $-function that forces one to choose a
better approach.

\subsection{The spin-phonon master equation}
\label{Sec-Spin-phonon-DME}

Using the expressions for the matrix elements of Eq.\ (\ref{VME}), for the
detailed transition probabilities of Eq.\ (\ref{GammatildeWW1}) one obtains
\begin{eqnarray}
\tilde{\Gamma}_{\mathcal{W};\mathcal{W},0_{i},+1_{\mathbf{k}}} &=&\frac{2\pi
}{N}\left| V_{\mathbf{k}}\right| ^{2}\mu _{i}\left( \nu _{\mathbf{k}%
}+1\right) \delta \left( \omega _{\mathbf{k}}-\omega _{0}\right)  \nonumber
\\
\tilde{\Gamma}_{\mathcal{W};\mathcal{W},1_{i},-1_{\mathbf{k}}} &=&\frac{2\pi
}{N}\left| V_{\mathbf{k}}\right| ^{2}\left( 1-\mu _{i}\right) \nu _{\mathbf{k%
}}\delta \left( \omega _{\mathbf{k}}-\omega _{0}\right) .
\label{Gammatildeik}
\end{eqnarray}
The master equation, Eq.\ (\ref{PWEq}), for the spin-phonon system \ then
becomes
\begin{eqnarray}
\frac{d}{dt}P_{\mathcal{W}} &=&\frac{2\pi }{N}\sum_{i}\mu _{i}\sum_{\mathbf{k%
}}\left| V_{\mathbf{k}}\right| ^{2}\left( \nu _{\mathbf{k}}+1\right) \delta
\left( \omega _{\mathbf{k}}-\omega _{0}\right)  \nonumber \\
&&\qquad \times \left( P_{\mathcal{W},0_{i},+1_{\mathbf{k}}}-P_{\mathcal{W}%
}\right)  \nonumber \\
&&{}+\frac{2\pi }{N}\sum_{i}\left( 1-\mu _{i}\right) \sum_{\mathbf{k}}\left|
V_{\mathbf{k}}\right| ^{2}\nu _{\mathbf{k}}\delta \left( \omega _{\mathbf{k}%
}-\omega _{0}\right)  \nonumber \\
&&\qquad \times \left( P_{\mathcal{W},1_{i},-1_{\mathbf{k}}}-P_{\mathcal{W}%
}\right) ,  \label{MasterSPExpl}
\end{eqnarray}
Let us now redefine $\nu _{\mathbf{k}}$ so that they refer to the state in
front of which they stand, instead of refferring to the state $\mathcal{W},$
as initialy defined. Then the master equation takes the form
\begin{eqnarray}
\frac{d}{dt}P_{\mathcal{W}} &=&\frac{2\pi }{N}\sum_{i}\sum_{\mathbf{k}%
}\left| V_{\mathbf{k}}\right| ^{2}\mu _{i}\delta \left( \omega _{\mathbf{k}%
}-\omega _{0}\right)  \nonumber \\
&&\qquad \times \left[ \nu _{\mathbf{k}}P_{\mathcal{W},0_{i},+1_{\mathbf{k}%
}}-\left( \nu _{\mathbf{k}}+1\right) P_{\mathcal{W}}\right]  \nonumber \\
&&{}+\frac{2\pi }{N}\sum_{i}\sum_{\mathbf{k}}\left| V_{\mathbf{k}}\right|
^{2}\left( 1-\mu _{i}\right) \delta \left( \omega _{\mathbf{k}}-\omega
_{0}\right)  \nonumber \\
&&\qquad \times \left[ \left( \nu _{\mathbf{k}}+1\right) P_{\mathcal{W}%
,1_{i},-1_{\mathbf{k}}}-\nu _{\mathbf{k}}P_{\mathcal{W}}\right] .
\label{MasterRedefined}
\end{eqnarray}

This master equation can be simplified by taking into account that all $%
N_{S} $ spins are identical and the state of the system depends on the
global spin excitation number
\begin{equation}
\mu =\sum_{i}\mu _{i},\qquad 0\leq \mu \leq N_{S}  \label{muDef}
\end{equation}
rather than on the particular spin configiration $\left\{ \mu _{i}\right\} $%
. Recalling the definition of the shortcuts for the quantum states in Eq.\ (%
\ref{Shortcuts}) one can replace $P_{\mathcal{W}}\Rightarrow P_{\left(
\mathcal{SP}\right) }\Rightarrow P_{\mu ,\left( \mathcal{P}\right) }$ and
rewrite Eq.\ (\ref{MasterRedefined}) as
\begin{eqnarray}
&&\frac{d}{dt}P_{\mu ,\left( \mathcal{P}\right) }=\frac{2\pi }{N}\sum_{%
\mathbf{k}}\left| V_{\mathbf{k}}\right| ^{2}\mu \delta \left( \omega _{%
\mathbf{k}}-\omega _{0}\right)  \nonumber \\
&&\qquad \times \left[ \nu _{\mathbf{k}}P_{\mu -1,\left( \mathcal{P},+1_{%
\mathbf{k}}\right) }-\left( \nu _{\mathbf{k}}+1\right) P_{\mu ,\left(
\mathcal{P}\right) }\right]  \nonumber \\
&&{}\qquad +\frac{2\pi }{N}\sum_{\mathbf{k}}\left| V_{\mathbf{k}}\right|
^{2}\left( N_{S}-\mu \right) \delta \left( \omega _{\mathbf{k}}-\omega
_{0}\right)  \nonumber \\
&&\qquad \times \left[ \left( \nu _{\mathbf{k}}+1\right) P_{\mu +1,\left(
\mathcal{P},-1_{\mathbf{k}}\right) }-\nu _{\mathbf{k}}P_{\mu ,\left(
\mathcal{P}\right) }\right] .  \label{MasterSPhmu}
\end{eqnarray}

\subsection{The observables}

The average excitation per spin is defined by
\begin{equation}
p=\frac{1}{N_{S}}\bar{\mu}=\frac{1}{N_{S}}\sum_{\mu =0}^{N_{S}}\frac{N_{S}!}{%
\mu !\left( N_{S}-\mu \right) !}\mu \prod_{\mathbf{k}}\sum_{\nu _{\mathbf{k}%
}=0}^{\infty }P_{\mu ,\left( \mathcal{P}\right) },  \label{ptilDef}
\end{equation}
where the binomial factor accounts for the degeneracy of the microscopic
spin states. One has $p=0$ if all spins are in the ground state and $p=1$ if
all spins are in the excited state. The average population of the phonon
mode ${\mathbf{k}}$ is defined in a similar way as
\begin{equation}
n_{\mathbf{k}}=\sum_{\mu =0}^{N_{S}}\frac{N_{S}!}{\mu !\left( N_{S}-\mu
\right) !}\sum_{\nu _{\mathbf{k}}=0}^{\infty }\nu _{\mathbf{k}}\prod_{%
\mathbf{q\neq k}}\sum_{\nu _{\mathbf{q}}=0}^{\infty }P_{\mu ,\left( \mathcal{%
P}\right) }.  \label{nkDef}
\end{equation}
It is never assumed that the spin and phonon parts of the density matrix
factorize.

It is convenient to introduce the distribution function for the spins
\begin{equation}
f_{\mu }\equiv \frac{N_{S}!}{\mu !\left( N_{S}-\mu \right) !}\prod_{\mathbf{k%
}}\sum_{\nu _{\mathbf{k}}=0}^{\infty }P_{\mu ,\left( \mathcal{P}\right) }
\label{fmuDef}
\end{equation}
that is normalized as $\sum_{\mu =0}^{N_{S}}f_{\mu }=1.$ Then Eq.\ (\ref
{ptilDef}) takes the form
\begin{equation}
p=\frac{1}{N_{S}}\bar{\mu}=\frac{1}{N_{S}}\sum_{\mu =0}^{N_{S}}\mu f_{\mu }.
\label{pfmu}
\end{equation}
For a macroscopic number of spins $N_{S}\gg 1$ the distribution function $%
f_{\mu }$ is sharply peaked at $\mu =\bar{\mu}.$ For instance, for
uncorrelated spins one has
\begin{equation}
f_{\mu }=\frac{N_{S}!}{\mu !\left( N_{S}-\mu \right) !}p^{\mu
}(1-p)^{N_{S}-\mu },  \label{Fmuuncorr}
\end{equation}
wherefrom follows $\bar{\mu}\cong \mu _{\max }\cong pN_{S}.$ Clearly $f_{\mu
}$ remains sharp if there is a correlation between the spins via the emitted
and absorbed phonons.

One can also introduce the conditional probability $n_{\mu ,\mathbf{k}}$ as
the total number of phonons in the $\mathbf{k}$-mode in the spin state $\mu $%
. It is defined by
\begin{equation}
n_{\mu ,\mathbf{k}}f_{\mu }=\frac{N_{S}!}{\mu !\left( N_{S}-\mu \right) !}%
\sum_{\nu _{\mathbf{k}}=0}^{\infty }\nu _{\mathbf{k}}\prod_{\mathbf{q\neq k}%
}\sum_{\nu _{\mathbf{q}}=0}^{\infty }P_{\mu ,\left( \mathcal{P}\right) },
\label{nmukDef}
\end{equation}
so that, evidently, $n_{\mathbf{k}}$ of Eq.\ (\ref{nkDef}) can be written as
\begin{equation}
n_{\mathbf{k}}=\sum_{\mu =0}^{N_{S}}n_{\mu ,\mathbf{k}}f_{\mu }.
\label{nkmuDef}
\end{equation}
Since $f_{\mu }$ is sharply peaked, we will use $n_{\mathbf{k}}\cong n_{\bar{%
\mu},\mathbf{k}}$ below.

\subsection{Kinetic equations for spins and phonons}

The time derivative of $p$ defined by Eq.\ (\ref{ptilDef}) can be calculated
with the help of master equation Eq.\ (\ref{MasterSPhmu}). The right-hand
side of the resulting equation can be simplified by introducing $f_{\mu }$
and $n_{\mu ,\mathbf{k}}$ with the help of Eq.\ (\ref{nmukDef}). After some
algebra one arrives at the equation
\begin{eqnarray}
&&\frac{dp}{dt}=\frac{\Gamma }{N_{S}}\sum_{\mu =0}^{N_{S}}\mu \left[ \left(
\mu +1\right) f_{\mu +1}-\mu f_{\mu }\right]  \nonumber \\
&&{}+\frac{2\pi }{NN_{S}}\sum_{\mathbf{k}}\left| V_{\mathbf{k}}\right|
^{2}\delta \left( \omega _{\mathbf{k}}-\omega _{0}\right) \sum_{\mu
=0}^{N_{S}}\mu  \nonumber \\
&&\times \left[ n_{\mu -1,\mathbf{k}}\left( N_{S}-\mu +1\right) f_{\mu
-1}-n_{\mu ,\mathbf{k}}\mu f_{\mu }\right]  \nonumber \\
&&{}+\frac{2\pi }{NN_{S}}\sum_{\mathbf{k}}\left| V_{\mathbf{k}}\right|
^{2}\delta \left( \omega _{\mathbf{k}}-\omega _{0}\right) \sum_{\mu
=0}^{N_{S}}\mu  \nonumber \\
&&\times \left[ n_{\mu +1,\mathbf{k}}\left( \mu +1\right) f_{\mu +1}-n_{\mu ,%
\mathbf{k}}\left( N_{S}-\mu \right) f_{\mu }\right] .  \label{pEqf}
\end{eqnarray}
The first line of this equation does not contain the phonon occupation
numbers and it describes the spontaneous emission of phonons. Here
\begin{equation}
\Gamma =\frac{2\pi }{N}\sum_{\mathbf{k}}\left| V_{\mathbf{k}}\right|
^{2}\delta \left( \omega _{\mathbf{k}}-\omega _{0}\right) =2\pi \left\langle
\left| V_{\mathbf{k}}\right| ^{2}\right\rangle \rho \left( \omega _{0}\right)
\label{GammaSingSp}
\end{equation}
is the single-spin decay rate,
\begin{equation}
\rho \left( \omega _{0}\right) =\frac{1}{N}\sum_{\mathbf{k}}\delta \left(
\omega _{\mathbf{k}}-\omega _{0}\right)  \label{rhoomega0Def}
\end{equation}
is the phonon density of states at the transition frequency of the spins,
and $\left\langle \left| V_{\mathbf{k}}\right| ^{2}\right\rangle $ is the
angular average
\begin{equation}
\left\langle \left| V_{\mathbf{k}}\right| ^{2}\right\rangle \equiv \int
\frac{dO_{\mathbf{k}}}{4\pi }\left| V_{\mathbf{k}}\right| ^{2}.
\label{G2AvrDef}
\end{equation}

Eq.\ (\ref{pEqf}) can be drastically simplified by shifting the $\mu $ index
under the sum over $\mu $, so that only $f_{\mu }$ enters. As a result one
obtains
\begin{eqnarray}
\frac{dp}{dt} &=&-\Gamma p+\frac{2\pi }{NN_{S}}\sum_{\mathbf{k}}\left| V_{%
\mathbf{k}}\right| ^{2}\delta \left( \omega _{\mathbf{k}}-\omega _{0}\right)
\nonumber \\
&&\qquad \times \sum_{\mu =0}^{N_{S}}f_{\mu }n_{\mu ,\mathbf{k}}\left(
N_{S}-2\mu \right) .
\end{eqnarray}
Here the sharpness of $f_{\mu },$ see Eq.\ (\ref{Fmuuncorr}) and the comment
below, leads to the final simplification. In the sum over $\mu $ one can
replace $\mu \Rightarrow \bar{\mu}=pN_{S}$ and $n_{\mu ,\mathbf{k}%
}\Rightarrow n_{\mathbf{k}},$ according to Eq.\ (\ref{nkmuDef}), and then
use the normalization condition for $f_{\mu }$, see Eq.\ (\ref{fmuDef}).
This leads to the final result
\begin{equation}
\frac{dp}{dt}=-\Gamma p+\left( 1-2p\right) \frac{2\pi }{N}\sum_{\mathbf{k}%
}\left| V_{\mathbf{k}}\right| ^{2}\delta \left( \omega _{\mathbf{k}}-\omega
_{0}\right) n_{\mathbf{k}}.  \label{pEqnk}
\end{equation}

The kinetic equation for the phonons can be derived in a similar way. The
result reads
\begin{equation}
\frac{dn_{\mathbf{k}}}{dt}=N_{S}\frac{2\pi }{N}\left| V_{\mathbf{k}}\right|
^{2}\delta \left( \omega _{\mathbf{k}}-\omega _{0}\right) \left[ p-(1-2p)n_{%
\mathbf{k}}\right] .  \label{nkEq}
\end{equation}
In fact, Eq.\ (\ref{nkEq}) could be guessed since, together with Eq.\ (\ref
{pEqnk}), it satisfies the excitation conservation
\begin{equation}
pN_{S}+\sum_{\mathbf{k}}n_{\mathbf{k}}=\mathrm{const}.  \label{excitconsev}
\end{equation}
A disappointing feature or Eqs.\ (\ref{pEqnk}) and (\ref{nkEq}) is that
powers of $\delta \left( \omega _{\mathbf{k}}-\omega _{0}\right) $ enter
both of them. If, say, there are no phonons in the initial state, then at
short times $n_{\mathbf{k}}$ will grow accordingly to Eq.\ (\ref{nkEq}) as $%
n_{\mathbf{k}}\varpropto t\delta \left( \omega _{\mathbf{k}}-\omega
_{0}\right) .$ Then multiplication of this by $\delta \left( \omega _{%
\mathbf{k}}-\omega _{0}\right) $ leads to a mathematically incorrect
expression. This problem was mentioned already in the discussion of the
applicability of the Pauli master equation at the end of Sec.\ \ref
{Sec-MasterEq}. The short-memory approach leading to the master equation is
inapplicable for the description of the decay of an initially prepared
state, if the reabsorption processes have to be taken into account, as in
the case of the phonon bottleneck.

\subsection{Ad hoc broadening of the $\protect\delta $-function}
\label{Sec-Broadening}

Instead of stepping back to correct the error made in the derivation of
Eqs.\ (\ref{pEqnk}) and (\ref{nkEq}), one can choose a cheap solution using
the regularization of the $\delta $-functions by ascribing them a finite
linewidth. One can, say, assume that the lineshapes are Lorentzian with a
linewidth $\Gamma ,$ such as radiational decay lineshape given by Eq.\ (\ref
{PWSE}). In this case the square of the $\delta $-function regularizes as
\begin{equation}
\delta ^{2}\left( \omega _{\mathbf{k}}-\omega _{0}\right) \Rightarrow \frac{1%
}{\pi \Gamma }\delta \left( \omega _{\mathbf{k}}-\omega _{0}\right) .
\end{equation}
Practically one can replace the energy $\delta $-function in Eq.\ (\ref
{pEqnk}) as
\begin{equation}
\delta \left( \omega _{\mathbf{k}}-\omega _{0}\right) \Rightarrow \frac{1}{%
\pi \Gamma },
\end{equation}
because there is one more delta function in $n_{\mathbf{k}}.$ In the
simplest case of $\left| V_{\mathbf{k}}\right| ^{2}=\left| V\right| ^{2}$
independent of the direction of $\mathbf{k}$ and no phonons in the initial
state one obtains
\begin{eqnarray}
\frac{dp}{dt} &=&-\Gamma p+\left( 1-2p\right) \frac{2\pi }{N}\frac{1}{\pi
\Gamma }\left| V\right| ^{2}\sum_{\mathbf{k}}n_{\mathbf{k}}  \nonumber \\
&=&-\Gamma p+\left( 1-2p\right) \frac{\Gamma }{N_{\Gamma }}N_{S}\left(
p_{0}-p\right) ,
\end{eqnarray}
where we used Eqs.\ (\ref{GammaSingSp}) and (\ref{NGammaDef}) to transform
\begin{equation}
\frac{2\left| V\right| ^{2}}{N\Gamma }=\frac{1}{N\pi \rho \left( \omega
_{0}\right) }=\frac{\Gamma }{N_{\Gamma }}  \label{GGamma}
\end{equation}
and used the conservation of the excitation, Eq.\ (\ref{excitconsev}), $p_{0}
$ being the spin excitation in the initial state. Finally one obtains
\begin{equation}
\frac{d}{dt}p=-\Gamma p+B\Gamma \left( p_{0}-p\right) (1-2p),
\end{equation}
where the bottleneck parameter $B$ is defined by Eq.\ (\ref{BDef}). This
equation is a particular case of the bottleneck equation that can be found
in the Abragam \& Bleany's book\cite{abrble70} \ and it is similar to all
other bottleneck equations \cite
{faustr61jpcs,scojef62pr,brywag67pr,pinfai90prb} published earlier and
later. Non of these publications provides a derivation of the bottleneck
equations so that it is impossible to judge whether these equations have
been written \emph{ad hoc} or derived in an incorrect way similar to that
described above.

\section{Dynamical equations for the phonon bottleneck}
\label{Sec-PB-Dynamical}

\subsection{Derivation and analysis of the equations}
\label{Sec-Derivation}

Having seen the origin of the breakdown of the kinetic description of the
PB, one can easily correct the error by stepping back to the master equation
with memory, Eq.\ (\ref{NEqMemory}). Then calculations similar to those of
Sec.\ \ref{Sec-Short-memory} yield the equations with memory for spins and
phonons
\begin{eqnarray}
\frac{dp}{dt} &=&-\frac{2}{N}\sum_{\mathbf{k}}\left| V_{\mathbf{k}}\right|
^{2}\int_{t_{0}}^{t}dt^{\prime }\cos \left[ \left( \omega _{\mathbf{k}%
}-\omega _{0}\right) \left( t-t^{\prime }\right) \right]  \nonumber \\
&&\qquad \times \left\{ p\left( t^{\prime }\right) +\left[ 2p\left(
t^{\prime }\right) -1\right] n_{\mathbf{k}}\left( t^{\prime }\right) \right\}
\label{pEqMem}
\end{eqnarray}
and
\begin{eqnarray}
\frac{dn_{\mathbf{k}}}{dt} &=&\frac{2N_{S}}{N}\left| V_{\mathbf{k}}\right|
^{2}\int_{t_{0}}^{t}dt^{\prime }\cos \left[ \left( \omega _{\mathbf{k}%
}-\omega _{0}\right) \left( t-t^{\prime }\right) \right]  \nonumber \\
&&\qquad \times \left\{ p\left( t^{\prime }\right) +\left[ 2p\left(
t^{\prime }\right) -1\right] n_{\mathbf{k}}\left( t^{\prime }\right)
\right\} ,  \label{nkEqMem}
\end{eqnarray}
instead of Eqs.\ (\ref{pEqnk}) and (\ref{nkEq}). As there is no more the
energy $\delta $-function in these equations, one has to deal with all
possible phonon modes. This means that in general the system of equations
above should be solved numerically.

As equations with memory are inconvenient for numerical solution, it is
better to remove the memory by introducing the additional dimensionless
variable
\begin{eqnarray}
r_{\mathbf{k}}(t) &=&\Gamma \int_{0}^{t}dt^{\prime }e^{i\left( \omega _{%
\mathbf{k}}-\omega _{0}\right) (t-t^{\prime })}  \nonumber \\
&&\qquad \times \left\{ p\left( t^{\prime }\right) +\left[ 2p\left(
t^{\prime }\right) -1\right] n_{\mathbf{k}}\left( t^{\prime }\right)
\right\} ,  \label{rkDef}
\end{eqnarray}
where $\Gamma $ is the spin-phonon decay rate given by Eq.\ (\ref
{GammaSingSp}). Evidently $r_{\mathbf{k}}(t)$ is related to the nondiagonal
element of the density matrix that was integrated out in Sec.\ \ref
{Sec-MasterEq}. One can see now that integrating out this nondiagonal
element was unneccesary. The same results could be obtained directly from
the Sch\"{o}dinger equation, Eq.\ (\ref{SEkmu}). The resulting system of
dynamical equations describing the direct spin-phonon processes has the form
\begin{eqnarray}
\frac{dp}{dt} &=&-\frac{1}{N}\sum_{\mathbf{k}}\frac{2\left| V_{\mathbf{k}%
}\right| ^{2}}{\Gamma }\func{Re}r_{\mathbf{k}}  \nonumber \\
\frac{dr_{\mathbf{k}}}{dt} &=&i\left( \omega _{\mathbf{k}}-\omega
_{0}\right) r_{\mathbf{k}}+\Gamma \left[ p+\left( 2p-1\right) n_{\mathbf{k}}%
\right]  \nonumber \\
\frac{dn_{\mathbf{k}}}{dt} &=&\frac{N_{S}}{N}\frac{2\left| V_{\mathbf{k}%
}\right| ^{2}}{\Gamma }\func{Re}r_{\mathbf{k}}.  \label{GEq}
\end{eqnarray}
Note that this system of equations is \emph{time-reversible}, as the
underlying SE, Eq.\ (\ref{SEkmu}). However, Eqs.\ (\ref{GEq}) are \emph{%
nonlinear} since they provide a reduced description of a many-body
quantum-mechanical system in terms of a few variables. We suppose that there
is no spin-phonon correlation in the initial state and use the initial
condition
\begin{equation}
r_{\mathbf{k}}(0)=0.  \label{rkIniCond}
\end{equation}
If $\left| V_{\mathbf{k}}\right| ^{2}=\left| V\right| ^{2}$ independently of
the direction of the emitted phonons, then with the help of Eqs.\ (\ref
{GammaSingSp}) and (\ref{BDef}) one can represent the coefficient in the
third equation as
\begin{equation}
\frac{N_{S}}{N}\frac{2\left| V\right| ^{2}}{\Gamma }=\Gamma B.
\label{GammaB}
\end{equation}
This gives an idea of the strength of the PB in Eqs.\ (\ref{GEq}). In the
case $B\ll 1$ and no phonons in the initial state, the generated phonon
populations $n_{\mathbf{k}}$ are small and can be neglected in the second of
Eqs.\ (\ref{GEq}). After that the latter can be integrated and the result
can be plugged into the first equation. Here one can make the short-memory
approximation $\func{Re}r_{\mathbf{k}}\cong \Gamma p\pi \delta \left( \omega
_{\mathbf{k}}-\omega _{0}\right) $ that is justified. This results into the
well-known pure decay equation $dp/dt=-\Gamma p.$

\begin{figure*}[t]
\includegraphics[angle=-90,width=8cm]{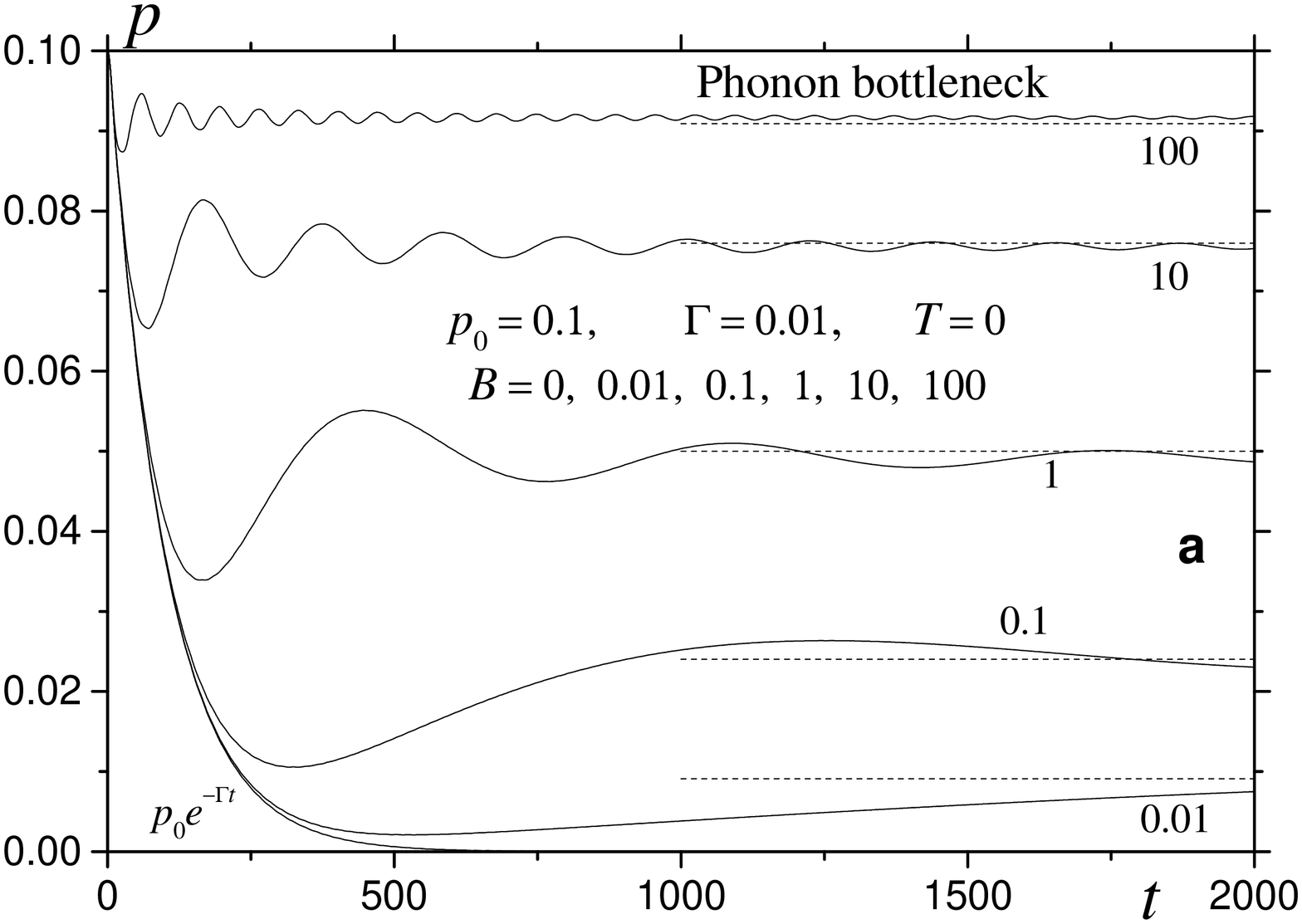} %
\includegraphics[angle=-90,width=8cm]{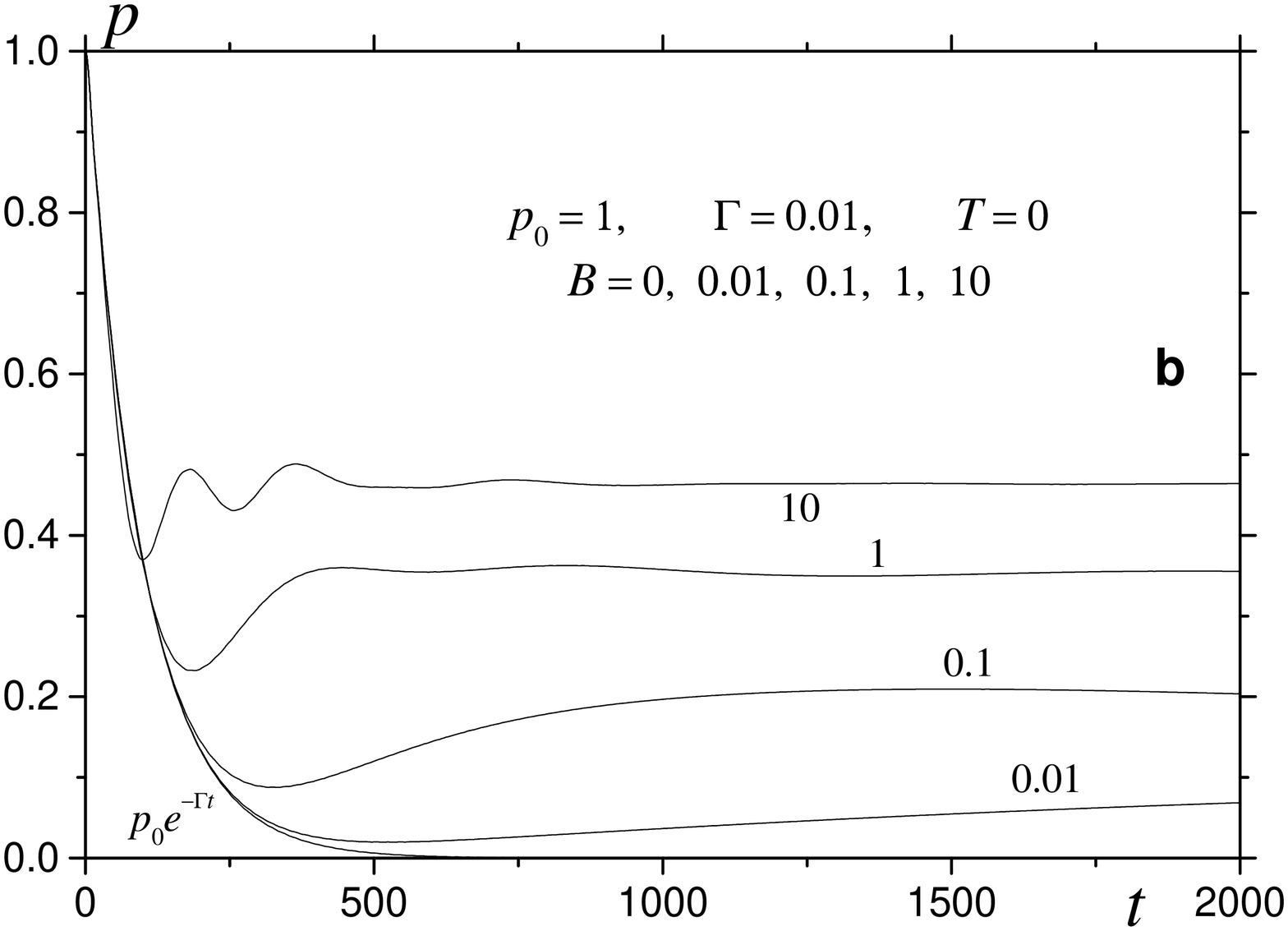}
\caption{Numerical solution of the dynamical equations for the
phonon bottleneck with nonrelaxing phonons at $T=0$ for different
values of the bottleneck parameter $B.$ (a) Low initial excitation
of spins, $p_{0}=0.1$. Asymptotes $p(\infty)$ given by Eq.\
(\protect\ref{pAsymptplateau}) with $n_0=0$ are
shown by dashed horizontal lines on the right. (b) High spin excitation, $%
p_{0}=1.$ In the case (b) the solution diverges for $B\gtrsim 100.$ The
bottleneck plateau $p(\infty)$ grows with $B$ but it should not exceed 1/2.}
\label{Fig-nonrelaxing-phonons}
\end{figure*}

At this point one can include the phonon dissipation into Eqs.\ (\ref{GEq})
in the simplest possible way, generalizing the method of Refs.\
\onlinecite
{faustr61jpcs,scojef62pr,brywag67pr,abrble70,pinfai90prb}:
\begin{eqnarray}
\frac{dp}{dt} &=&-\frac{1}{N}\sum_{\mathbf{k}}\frac{2\left| V_{\mathbf{k}%
}\right| ^{2}}{\Gamma }\func{Re}r_{\mathbf{k}}  \nonumber \\
\frac{dr_{\mathbf{k}}}{dt} &=&\left[ i\left( \omega _{\mathbf{k}}-\omega
_{0}\right) -\frac{1}{2}\Gamma _{\mathrm{ph}}\right] r_{\mathbf{k}}+\Gamma %
\left[ p+\left( 2p-1\right) n_{\mathbf{k}}\right]  \nonumber \\
\frac{dn_{\mathbf{k}}}{dt} &=&\Gamma _{\mathrm{ph}}\left( n_{\mathrm{eq}}-n_{%
\mathbf{k}}\right) +\frac{N_{S}}{N}\frac{2\left| V_{\mathbf{k}}\right| ^{2}}{%
\Gamma }\func{Re}r_{\mathbf{k}}.  \label{GEqsPhDiss}
\end{eqnarray}
Here $\Gamma _{\mathrm{ph}}$ is the phonon damping and $n_{\mathrm{eq}}$ is
the thermally equilibrium phonon population near the resonance,
\begin{equation}
n_{\mathrm{eq}}=\frac{1}{\exp \left( \frac{\hbar \omega _{0}}{k_{B}T}\right)
-1}.  \label{nkEqui}
\end{equation}
At equilibrium one has $n_{\mathbf{k}}=n_{\mathrm{eq}},$ $r_{\mathbf{k}}=0,$
and $p$ is obtained from the equation $p+\left( 2p-1\right) n_{\mathbf{k}}=0$
that yields
\begin{equation}
p_{\mathrm{eq}}=\frac{n_{\mathrm{eq}}}{1+2n_{\mathrm{eq}}}=\frac{1}{\exp
\left( \frac{\hbar \omega _{0}}{k_{B}T}\right) +1},  \label{pEqui}
\end{equation}
an expected result.

In the case $\Gamma _{\mathrm{ph}}\gg \Gamma $ the variables $r_{\mathbf{k}}$
and $n_{\mathbf{k}}$ can be expected to relax much faster than $p.$ Thus
after some time they should adiabatically adjust to the instantaneous value
of $p.$ The same will happen in the case of strong bottleneck at
asymptotically large times. Neglecting $\dot{r}_{\mathbf{k}}$ and $\dot{n}_{%
\mathbf{k}}$ in Eq. (\ref{GEqsPhDiss}) one obtains
\begin{equation}
n_{\mathbf{k}}=n_{\mathrm{eq}}+\frac{\frac{N_{S}}{N}\left| V_{\mathbf{k}%
}\right| ^{2}\left[ p+\left( 2p-1\right) n_{\mathrm{eq}}\right] }{\left(
\omega _{\mathbf{k}}-\omega _{0}\right) ^{2}+\frac{\Gamma _{\mathrm{ph}}^{2}%
}{4}-\frac{N_{S}}{N}\left| V_{\mathbf{k}}\right| ^{2}\left( 2p-1\right) }
\label{nkAdjusted}
\end{equation}
for the adjusted value of $n_{\mathbf{k}}.$ The second nonequilibrium term
here is small and it has a Lorentz line shape with the phonon line width $%
\Gamma _{\mathrm{ph}}$ if the last term in the denominator can be neglected.
Note that for large $\Gamma _{\mathrm{ph}}$ the distribution of emitted
phonons becomes a smooth enough function to make the short-memory Pauli
master equation applicable [see discussion below Eq. (\ref{NEqMemory})]. In
this case, however, the problem trivializes and the bottleneck disappears.
One can see that the line width of emitted phonons narrows in the case of
the inverse spin population, $p>1/2.$ The reason is the stimulated phonon
emission acting as phonon pumping that competes with their natural damping $%
\Gamma _{\mathrm{ph}}.$ The strength of the effect depends on the bottleneck
parameter $B,$ as follows from Eq. (\ref{GammaB}).

Similarly to Eq. (\ref{nkAdjusted}), one can find the adjusted value $\func{%
Re}r_{\mathbf{k}}$ and plug it into the first of Eq. (\ref{GEqsPhDiss}). If $%
\left| V_{\mathbf{k}}\right| ^{2}=\left| V\right| ^{2}$ independently of the
direction of the emitted phonons, integration over $\mathbf{k}$ yields the
equation

\begin{equation}
\frac{dp}{dt}=-\Gamma ^{\ast }(p)\left[ p+\left( 2p-1\right) n_{\mathrm{eq}}%
\right] ,  \label{pEq}
\end{equation}
where
\begin{equation}
\Gamma ^{\ast }(p)=\Gamma \left[ 1-\frac{2B\left( 1-2p\right) \Gamma ^{2}}{%
\left( \Gamma _{\mathrm{ph}}+\Gamma _{\mathrm{ph}}^{\ast }(p)\right) \Gamma
_{\mathrm{ph}}^{\ast }(p)}\right]   \label{Gammastarp}
\end{equation}
and
\begin{equation}
\Gamma _{\mathrm{ph}}^{\ast }(p)\equiv \sqrt{\Gamma _{\mathrm{ph}%
}^{2}+2B\left( 1-2p\right) \Gamma ^{2}}.  \label{Gammaphstarp}
\end{equation}
One can see that the condition
\begin{equation}
\Gamma _{\mathrm{ph}}\gtrsim \sqrt{B}\Gamma   \label{GammaphStrong}
\end{equation}
is needed to ensure $\Gamma ^{\ast }(p)\cong \Gamma $ and thus to suppress
the bottleneck. For $B\gg 1,$ this is a stronger condition than the
first-glance expectation $\Gamma _{\mathrm{ph}}\gtrsim \Gamma .$ If Eq. (\ref
{GammaphStrong}) is satisfied, then the well-known solution of this equation
\begin{equation}
p(t)=p_{\mathrm{eq}}+(p_{0}-p_{\mathrm{eq}})e^{-\Gamma _{T}t},\qquad \Gamma
_{T}=\Gamma (1+2n_{\mathrm{eq}})  \label{ptGenRelax}
\end{equation}
is recovered. In the case of strong bottleneck, opposite to Eq. (\ref
{GammaphStrong}), simplification of Eq. (\ref{Gammastarp}) yields
\begin{equation}
\Gamma ^{\ast }(p)\cong \frac{\Gamma _{\mathrm{ph}}}{\sqrt{2B\left(
1-2p\right) }}  \label{GammastarpAsymp}
\end{equation}
for $p<1/2$. One can see that for strong bottleneck the effective spin
relaxation rate does not depend on $\Gamma $ at all. This means that the
transfer of energy from the resonant phonons to the rest of the phonon bath
or elsewhere is really the bottleneck of the whole process. On the other
hand, the effective spin relaxation rate is not just $\Gamma _{\mathrm{ph}},$
as one could assume.

It should be stressed that in the case $B\gg 1$ and Eq. (\ref{GammaphStrong}%
) not satisfied, Eq. (\ref{pEq}) becomes a reasonable approximation at
asymptotically large times only. At small and intermediated times the full
solution of Eq. (\ref{GEqsPhDiss}) shows oscillations of $p$ similar to
probability oscillations in the soluition for two coupled quantum-mechanical
states. \ The heuristic condition $\Gamma _{\mathrm{ph}}\gtrsim \Gamma $
mentioned at the beginning of this analysis is insufficient to establish
adiabatic adjustment of $r_{\mathbf{k}}$ and $n_{\mathbf{k}}$ to the
instantaneous value of $p$ because of the large term $\sim B$ in the third
of Eq. (\ref{GEqsPhDiss}).

\begin{figure}[t]
\includegraphics[angle=-90,width=8cm]{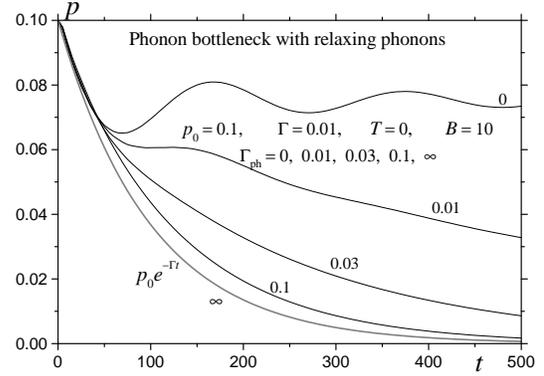}
\caption{Numerical solution for the spin excitation $p(t)$ of the
dynamical equations with relaxing phonons, Eq.\
(\protect\ref{GEqsPhDiss}), at $T=0$ for $B=10$ and different
values of the phonon relaxation rate $\Gamma _{\mathrm{ph}}$. Note
that $\Gamma _{\mathrm{ph}}\gg \Gamma $ is needed to suppress the
bottleneck, if $B\gg 1$.} \label{Fig-relaxing-phonons-T0}
\end{figure}

\begin{figure}[t]
\includegraphics[angle=-90,width=8cm]{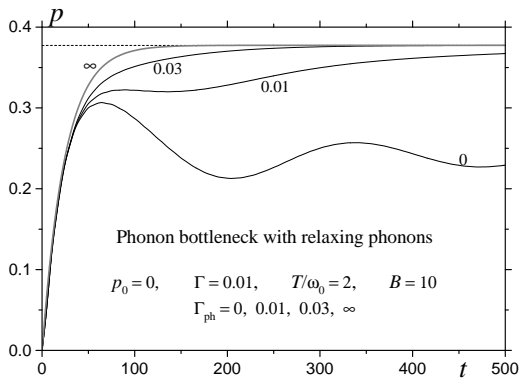}
\caption{Numerical solution for $p(t)$ with relaxing phonons and $%
k_{B}T/(\hbar \protect\omega _{0})=2$. Here the spins are initially in the
ground state, $p_{0}=0$, and they absorb the energy from phonons. For $%
\Gamma _{\mathrm{ph}}=0$, depletion of the phonons prevents reaching the
thermal equilibrium value of $p$ (the dotted horizontal line). }
\label{Fig-relaxing-phonons-T1}
\end{figure}

\begin{figure*}[t]
\includegraphics[angle=-90,width=8cm]{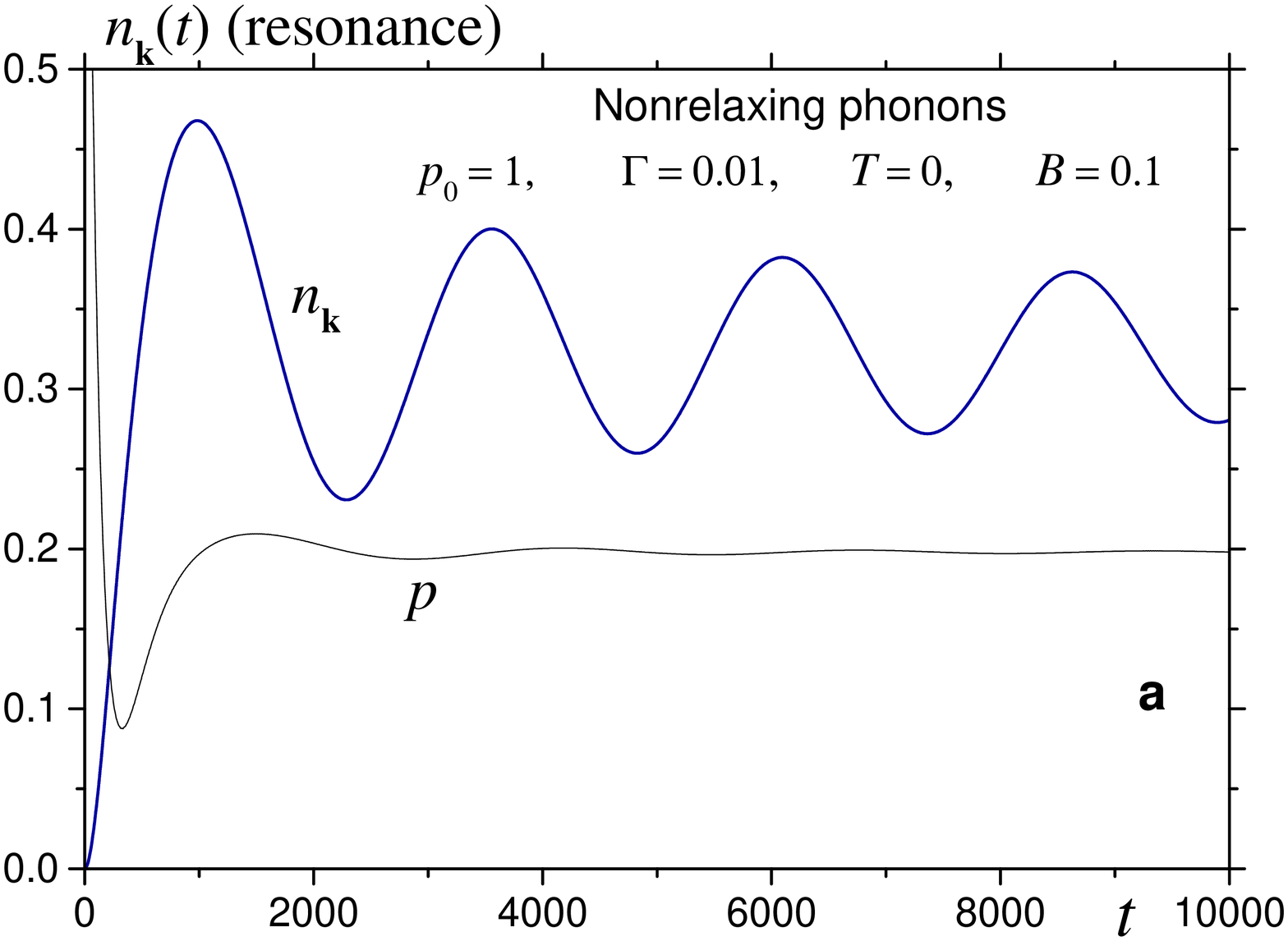} %
\includegraphics[angle=-90,width=8cm]{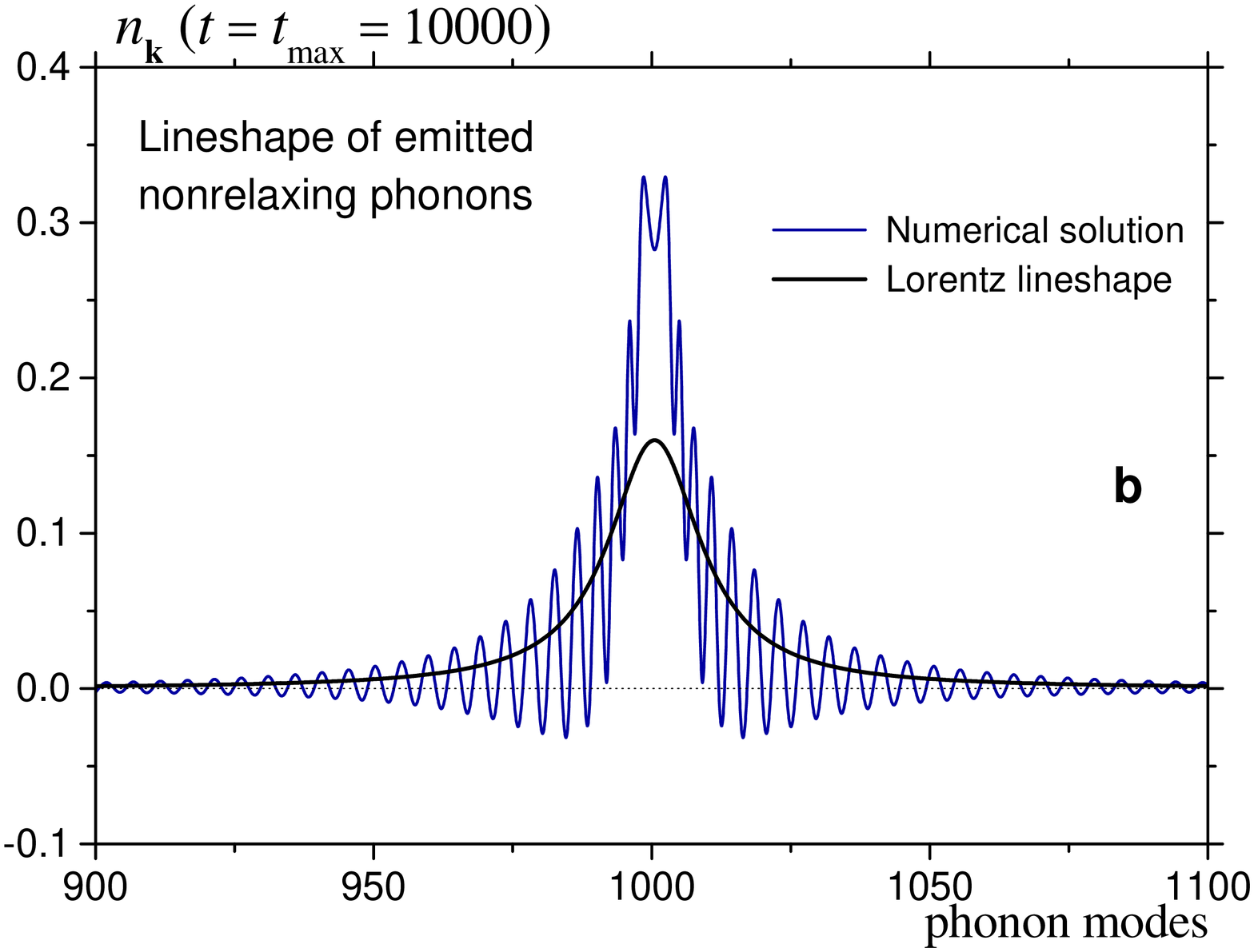}
\caption{Emitted phonons in the absence of the phonon relaxation
at $T=0$ for a moderate bottleneck, $B=0.1$. (a) Time dependence
$n_{\mathbf{k}}(t)$ at resonance remains oscillating long after
the spin excitation $p(t)$ reached the bottleneck plateau; (b)
Lineshape of the emitted phonons (at the maximal time) is narrow
but not everywhere positive, that indicates that Eq.\
(\protect\ref{GEq}) needs improving.} \label{Fig-nk}
\end{figure*}

\subsection{Numerical solution of the dynamical spin-phonon equations}
\label{Sec-Numerical}

It is not difficult to solve Eqs.\ (\ref{GEq}) and (\ref{GEqsPhDiss})
numerically. The results of the numerical solution of Eqs.\ (\ref{GEq}) for
macroscopic samples and no phonons in the initial state (i.e., $T=0)$ are
shown in Fig.\ \ref{Fig-nonrelaxing-phonons}. The direct spin-phonon
relaxation rate was taken to be $\Gamma =0.01$ and the phonon cut-off
frequency $\omega _{\max }\sim 1,$ so that the short-memory approximation is
valid in the absence of the bottleneck. For simplicity, the phonon modes
have been discretized equidistantly, ensuring $N_{\Gamma }\gg 1$. Here and
in all other numerical calculations we have set $\left| V_{\mathbf{k}%
}\right| ^{2}=\left| V\right| ^{2}$ for simplicity, to use Eq.\ (\ref{GammaB}%
). One can see that for $B=0$ the solution for $p(t)$ \ is the pure-decay
exponential, $p=p_{0}e^{-\Gamma t}$ practically in the whole time domain.
The initial quadratic dependence of $p(t)$ stemming from the time
reversibility of Eq.\ (\ref{GEq}) is confined to very short times of order $%
1/\omega _{\max }$ and it is not seen in Fig.\ \ref{Fig-nonrelaxing-phonons}%
. The results in Fig.\ \ref{Fig-nonrelaxing-phonons}a for the initial spin
excitation $p_{0}=0.1$ show \emph{oscillating} approaching a plateau that
becomes higher with increasing $B.$ These oscillations is a memory effect
that is absent in the earlier theories of the PB.  \cite
{faustr61jpcs,scojef62pr,brywag67pr,abrble70,pinfai90prb} For low spin
excitations, $p\ll 1,$ the bottleneck equations can be solved analytically
(see below). The analytically obtained asymptotic values of $p$ are shown in
Fig.\ \ref{Fig-nonrelaxing-phonons}a by dashed horizontal lines on the right.

Numerical results for the initially fully excited spin system, $p_0=1,$ are
represented in Fig.\ \ref{Fig-nonrelaxing-phonons}b. The results are
qualitatively similar to those in the case of the low spin excitation, Fig.\
\ref{Fig-nonrelaxing-phonons}a, if the bottleneck parameter $B$ is not too
high. However, for $B=100$ the solution of Eq.\ (\ref{GEq}) shows an
unphysical divergence with $p$ going to infinity. This indicates that our
bottleneck equations are not accurate enough to describe the high-excitation
strong-bottleneck case. High excitation means the inverted population in the
initial spin state, $p_{0}>1/2,$ that formally corresponds to the \emph{%
negative} spin temperature. Asymptotically the spin temperature should
equilibrate with the temperature of the resonant phonon group. As the phonon
energies are not bounded from above, phonons cannot have a negative
temperature. Thus also spins cannot have a negative temperature in the
asymptotic state. One can expect that for the initial spin inversion and
huge values of $B$ the asymptotic common spin-phonon temperature will be
just very high, i.e., the asymptotic value of $p$ will be close to 1/2. This
means that the asymptotic value of $p$ should saturate at 1/2 with
increasing $B.$ This tendency is seen in Fig.\ \ref{Fig-nonrelaxing-phonons}%
b, only instead of the saturation at $p=1/2$ one obtains a divergence.

Figs. \ref{Fig-relaxing-phonons-T0} and \ref{Fig-relaxing-phonons-T1} show
the numerical solution of Eqs.\ (\ref{GEqsPhDiss}) for $B=10$ and different
values of the phonon relaxation rate $\Gamma _{\mathrm{ph}}$. In Fig.\ \ref
{Fig-relaxing-phonons-T0} the temperature of the phonon bath is $T=0$ and
the starting spin excitation is $p_{0}=0.1.$ One can see that for all $%
\Gamma _{\mathrm{ph}}>0$ the spin excitation $p$ relaxes to zero. The
effective spin relaxation rate increases with $\Gamma _{\mathrm{ph}}$ and
asymptotically approaches $\Gamma ,$ as the PB is gradually suppressed by
the phonon relaxation. However, the values of $\Gamma _{\mathrm{ph}}$
neeeded to achieve the effective rate close to $\Gamma $ are substantially
greater than $\Gamma ,$ in accordance with the estimation in Eq.\ (\ref
{GammaphStrong}).

Fig.\ \ref{Fig-relaxing-phonons-T1} shows the PB with the opposite direction
of the relaxation, also with $B=10$ for different values of $\Gamma _{%
\mathrm{ph}}$. Here $p_{0}=0$ and the temperature is nonzero, $k_{B}T/(\hbar
\omega _{0})=2.$ Thus the spins absorb the energy from the phonon bath. If $%
\Gamma _{\mathrm{ph}}=0$ and $B\gtrsim 1$, there are too little phonons in
resonance with the spins, so that the resonant phonons get depleted and the
spins cannot reach the equilibrium value $p_{\mathrm{eq}}$ given by Eq.\ (%
\ref{pEqui}). However, for all $\Gamma _{\mathrm{ph}}>0$ the phonons are
replenished and the system reaches the equilibrium. Again, for $\Gamma _{%
\mathrm{ph}}$ satisfying the strong inequality of Eq.\ (\ref{GammaphStrong}%
), the spin relaxation curve acquires its standard form without the PB,
given by Eq.\ (\ref{ptGenRelax}).

Studying the time evolution of the lineshape of emitted phonons, in the
absence of the phonon relaxation, gives additional insights into the
dynamics described by Eq.\ (\ref{GEq}). Fig.\ \ref{Fig-nk} shows the
numerical results obtained with $p_{0}=1$ and $B=0.1.$ The time dependence
of $n_{\mathbf{k}}$ at resonance in Fig.\ \ref{Fig-nk}a remains oscillating
at very large times, where $p(t)$ is already a constant. The lineshape $n_{%
\mathbf{k}}$ at the maximal time of the calculation plotted vs the
phonon-mode number (1000 corresponds to the resonance) is shown in Fig.\ \ref
{Fig-nk}b. It is strongly oscillating, and $n_{\mathbf{k}}$ becomes negative
for some modes at some moments of time. That is, the lineshape oscillates
both as the function of the phonon mode number and the time, as Fig.\ \ref
{Fig-nk}a indicates. At large times, when the excitation of the spins has
already reached its asymptotic value, the energy is still being exchanged
between different phonon modes of the resonant group. Even for smaller
values of $B,$ the lineshape does not approach the radiational lineshape of
Eq.\ (\ref{PWSE}) that is also shown in Fig.\ \ref{Fig-nk}b. This implies
that Eq.\ (\ref{GEq}) is still not accurate enough to correctly describe the
lineshape of the emitted phonons. Still it yields some lineshape, whereas
earlier theories \cite
{faustr61jpcs,scojef62pr,brywag67pr,abrble70,pinfai90prb} did not consider
the lineshape at all.

\subsection{Bottleneck at low excitation}
\label{Sec-LowExc}

In the case of low excitation of the spins throughout the process, $p(t)\ll
1,$ Eq.\ (\ref{GEqsPhDiss}) can be linearized by setting $p+\left(
2p-1\right) n_{\mathbf{k}}\Rightarrow p-n_{\mathbf{k}}$ and then it can be
solved by the Fourier transformation. We define the functions on the whole
time interval ($-\infty ,\infty $) and set them to zero at $t<0.$ Then one
has to introduce the initial-condition terms $p_{0}\delta (t)$ and $%
n_{0}\delta (t)$ into the first and third equations. Note that the initial
phonon occupation can differ from the equilibrium value $n_{\mathrm{eq}}.$
Fourier transforms are defined by
\begin{equation}
p(t)=\int_{-\infty }^{\infty }\frac{d\omega }{2\pi }\tilde{p}(\omega
)e^{-i\omega t},\qquad \tilde{p}(\omega )=\int_{-\infty }^{\infty
}dtp(t)e^{i\omega t},
\end{equation}
etc. With
\begin{equation}
r_{\mathbf{k}}=\xi _{\mathbf{k}}+i\psi _{\mathbf{k}}  \label{rkxipsiDef}
\end{equation}
Eq.\ (\ref{GEqsPhDiss}) can be put into the form
\begin{eqnarray}
\frac{dp}{dt} &=&p_{0}\delta (t)-\frac{1}{N}\sum_{\mathbf{k}}\frac{2\left|
V_{\mathbf{k}}\right| ^{2}}{\Gamma }\xi _{\mathbf{k}}  \nonumber \\
\frac{dn_{\mathbf{k}}}{dt} &=&n_{0}\delta (t)+\Gamma _{\mathrm{ph}}\left[ n_{%
\mathrm{eq}}\theta \left( t\right) -n_{\mathbf{k}}\right] +\frac{N_{S}}{N}%
\frac{2\left| V_{\mathbf{k}}\right| ^{2}}{\Gamma }\xi _{\mathbf{k}}
\nonumber \\
\frac{d\xi _{\mathbf{k}}}{dt} &=&-\left( \omega _{\mathbf{k}}-\omega
_{0}\right) \psi _{\mathbf{k}}-\frac{1}{2}\Gamma _{\mathrm{ph}}\xi _{\mathbf{%
k}}+\Gamma \left[ p-n_{\mathbf{k}}\right]  \nonumber \\
\frac{d\psi _{\mathbf{k}}}{dt} &=&\left( \omega _{\mathbf{k}}-\omega
_{0}\right) \xi _{\mathbf{k}}-\frac{1}{2}\Gamma _{\mathrm{ph}}\psi _{\mathbf{%
k}}.  \label{xipsiEq}
\end{eqnarray}
After the Fourier transformation one obtains a system of linear algebraic
equations
\begin{eqnarray}
-i\omega \tilde{p} &=&p_{0}-\frac{1}{N}\sum_{\mathbf{k}}\frac{2\left| V_{%
\mathbf{k}}\right| ^{2}}{\Gamma }\tilde{\xi}_{\mathbf{k}}  \nonumber \\
-i\omega \tilde{n}_{\mathbf{k}} &=&n_{0}+\Gamma _{\mathrm{ph}}\left\{ \frac{%
n_{\mathrm{eq}}}{-i\omega +0}-\tilde{n}_{\mathbf{k}}\right\} +\frac{N_{S}}{N}%
\frac{2\left| V_{\mathbf{k}}\right| ^{2}}{\Gamma }\tilde{\xi}_{\mathbf{k}}
\nonumber \\
-i\omega \tilde{\xi}_{\mathbf{k}} &=&-\left( \omega _{\mathbf{k}}-\omega
_{0}\right) \tilde{\psi}_{\mathbf{k}}-\frac{1}{2}\Gamma _{\mathrm{ph}}\tilde{%
\xi}_{\mathbf{k}}+\Gamma \left[ \tilde{p}-\tilde{n}_{\mathbf{k}}\right]
\nonumber \\
-i\omega \tilde{\psi}_{\mathbf{k}} &=&\left( \omega _{\mathbf{k}}-\omega
_{0}\right) \tilde{\xi}_{\mathbf{k}}-\frac{1}{2}\Gamma _{\mathrm{ph}}\tilde{%
\psi}_{\mathbf{k}}.  \label{xipsiEqFourier}
\end{eqnarray}
Here one eliminates $\tilde{n}_{\mathbf{k}}$ and $\tilde{\psi}_{\mathbf{k}}$
and expresses $\tilde{\xi}_{\mathbf{k}}$ through $\tilde{p}.$ Then plugging $%
\tilde{\xi}_{\mathbf{k}}$ into the first equation yields an isolated
equation for $\tilde{p}$ that can be solved. In the resulting expression for
$\tilde{p}$ one sums over $\mathbf{k.}$ Making then the inverse Fourier
transformation one obtains, in particular, the asymptotic value
\begin{equation}
p(\infty )=\frac{p_{0}\sqrt{B}+n_{0}}{1+\sqrt{B}}  \label{pAsymptplateau}
\end{equation}
in the case $\Gamma _{\mathrm{ph}}=0,$ describing the phonon-bottleneck
plateau. Also for $\Gamma _{\mathrm{ph}}=0$ one obtains
\begin{equation}
\xi _{\mathbf{k}}(t)=\left( p_{0}-n_{0}\right) \frac{\sqrt{B}}{1+\sqrt{B}}%
\frac{\sin \left( \Omega _{\mathbf{k}}t\right) }{\Omega _{\mathbf{k}}},
\label{ksiktSol}
\end{equation}
where
\begin{equation}
\Omega _{\mathbf{k}}\equiv \sqrt{\left( \omega _{\mathbf{k}}-\omega
_{0}\right) ^{2}+B\Gamma ^{2}}.  \label{OmegakpsmallDef}
\end{equation}
This result features the avoided level crossing between the energy levels of
the spins and phonons, as $\xi _{\mathbf{k}}(t)$ that describes the
dynamical spin-phonon correlations continues to oscillate even at resonance,
$\omega _{\mathbf{k}}-\omega _{0}.$ The splitting $\Delta =B\Gamma ^{2}$ in $%
\Omega _{\mathbf{k}}$ is proportional to the concentration of spins via $B$
given by Eq.\ (\ref{BDef}). One can see from Eq.\ (\ref{GammaB}) that in the
case of nondiluted spins, $N_{S}=N,$ one has $\Delta =B\Gamma ^{2}=2\left|
V\right| ^{2}.$ This result is similar to that of Ref.\ %
\onlinecite{jacste63pr}.

For the model with relaxing phonons, in the strong-bottleneck limit $B\gg 1$
one obtains the following asymptotic ($\Gamma _{\mathrm{ph}}t\gg 1$)
relaxation law
\begin{equation}
p(t)=\left( p_{0}+\frac{\widetilde{\Gamma }}{\Gamma _{\mathrm{ph}}}%
n_{0}\right) e^{-\widetilde{\Gamma }t}+\left( 1-e^{-\widetilde{\Gamma }%
t}\right) n_{\mathrm{eq}},
\end{equation}
where
\begin{equation}
\widetilde{\Gamma }\equiv \frac{\Gamma \Gamma _{\mathrm{ph}}}{\sqrt{\Gamma _{%
\mathrm{ph}}^{2}+2B\Gamma ^{2}}}
\end{equation}
is the effective relaxation rate of the spin-phonon system in the case of
strong bottleneck, $B\gg 1$. One can see that for $\Gamma _{\mathrm{ph}}\ll
\sqrt{B}\Gamma $ the effective rate is $\widetilde{\Gamma }\cong \Gamma _{%
\mathrm{ph}}/\sqrt{2B}\ll \Gamma _{\mathrm{ph}}.$ The same result follows
from  Eq. (\ref{GammastarpAsymp}) in the limit $p\ll 1.$ For $B\ll 1$ in
Eq.\ (\ref{GammaphStrong}) the effective rate becomes $\widetilde{\Gamma }%
\cong \Gamma $ and the bottleneck disappears.

\subsection{Asymptotic stability of the bottleneck equations}
\label{Sec-AsymtStab}

With the methods of the previous section it is possible to analyze the
asymptotic stability of the bottleneck equations with undamped phonons, Eq.\
(\ref{GEq}). Asymptotically $p=\mathrm{const},$ so that one can substitute
it into Eq.\ (\ref{xipsiEq}) with $\Gamma _{\mathrm{ph}}=0.$ Assuming, for
simplicity, $\left| V_{\mathbf{k}}\right| ^{2}=\left| V\right| ^{2}=\mathrm{%
const}$ and introducing a new variable $\phi _{\mathbf{k}}$ as
\begin{equation}
\psi _{\mathbf{k}}=\phi _{\mathbf{k}}+\frac{\Gamma p}{\omega _{\mathbf{k}%
}-\omega _{0}},
\end{equation}
one obtains the system of linear ordinary differential equations
\begin{eqnarray}
\dot \xi _{\mathbf{k}} &=&-\left( \omega _{\mathbf{k}}-\omega _{0}\right)
\phi _{\mathbf{k}}+\left( 2p-1\right) \Gamma n_{\mathbf{k}}  \nonumber \\
\dot \phi _{\mathbf{k}} &=&\left( \omega _{\mathbf{k}}-\omega _{0}\right)
\xi _{\mathbf{k}}  \nonumber \\
\dot n_{\mathbf{k}} &=&B\Gamma \xi _{\mathbf{k}}.
\end{eqnarray}
The eigenfrequencies of this system of equations are $\Omega _{\mathbf{k}}=0$
and
\begin{equation}
\Omega _{\mathbf{k}}=\pm \sqrt{\left( \omega _{\mathbf{k}}-\omega
_{0}\right) ^{2}+B\Gamma ^{2}(1-2p)},
\end{equation}
c.f. Eq.\ (\ref{OmegakpsmallDef}). For the inverse population, $p>1/2,$ the
eigenfrequency $\Omega _{\mathbf{k}}$ becomes imaginary in the frequency
region near the resonance. This leads to an instability that is seen in the
numerical results as positive divergence of $p$ and thus negative divergence
of $n_{\mathbf{k}}$ near the resonance. The problem is that for the
inversely-populated initial spin states, $p_{0}>1/2$ and $B\gg 1$ the
asymptotic value of $p$ is close to 1/2, as discussed above. Then even small
inaccuracies of the dynamic equations can render $p$ slightly exceeding 1/2
that leads to the divergency. The fact that for $p_{0}=1$ the instability
happens only starting from $B\sim 100$ (see Fig.\ \ref
{Fig-nonrelaxing-phonons}b) where $p(\infty )$ should be already very close
to 1/2, indicates that the instability is driven by the inaccuracies of Eq.\
(\ref{GEq}).

\section{Discussion}
\label{Sec-Summary}

In the main part of this paper it was shown that the dynamics of two-level
systems (spins) interacting with a continuum of resonant phonons via direct
emisson and absorption processes, including the phonon-bottleneck effect, is
much more complicated than generally accepted. Since the emitted/absorbed
phonon packet is narrow in the energy space, the usual \emph{kinetic}
description based on the short-memory Pauli master equation is invalid.
Taking into account memory amounts to stepping back to a \emph{dynamic}
description including nondiagonal elements of the density matrix. This cures
the problem of the powers of the energy $\delta $-functions in the standard
formalism while being capable of describing the PB. The resulting system of
dynamical equations, Eq.\ (\ref{GEq}), can be enhanced to include the
relaxation of phonons in a simple way, resulting in Eq.\ (\ref{GEqsPhDiss}).
For the low spin excitation, $p\ll 1,$ these equations linearize and can be
solved analytically. In a number of particular cases, such as small
bottleneck parameter $B$ of Eq.\ (\ref{BDef}) or fast phonon relaxation rate
$\Gamma _{\mathrm{ph}}$, see Eq.\ (\ref{GammaphStrong}), simple results
without a bottleneck are reproduced.

The approach formulated here is not a final solution of the long-standing PB
problem but rather a next step in improving the existing theories. A
deficiency of the present approach is its inability to provide an accurate
lineshape for emitted phonons, even the well-known Lorentzian shape of Eq.\ (%
\ref{BDef}) for the weak bottleneck, $B\ll 1.$ This should be the reason for
the instability of Eq.\ (\ref{GEq}) in the case of the strong bottleneck, $%
B\gg 1,$ with spin inversion in the initial state. As the dynamical
bottleneck equations have been obtained within the same method as the Pauli
master equation, that is, by cutting the infinite chain of coupled equations
for the nondiagonal elements of the density matrix at the lowest possible
level, an evident idea of improvement is to take into account the
next-generation nondiagonal matrix elements. This would result in a more
complicated formalism, however, that deserves to be dealt with in a separate
work.

In real systems the spin transition frequency $\omega _{0}$ is
inhomogeneously broadened. That is, there are different transition
frequencies $\omega _{i}$ for different spins, distributed around the
average value $\omega _{0}$ with a width $\Delta \omega _{0}\gg \Gamma .$
Obviously this leads to the reduction of the PB effect since more phonon
modes can exchange excitation with the spin system, or, in other words, less
spins can exchange excitation with a given phonon mode. Thus one can expect
that the situation will depend on the effective bottleneck parameter $\tilde{%
B}\sim (\Gamma /\Delta \omega _{0})B\ll B.$

Incorporating inhomogeneous broadening in the theory seems to be
straightforward but it would result in a serious complication of the
formalism. The problem is that spins cannot be described by a single
variable $p$ any longer. Spin with each particular $\omega _{i}$ should
exhibit their own dynamics, so that one has to search for the distribution $%
p(\omega _{i})$. It is not clear if any analytical solutions for the
corresponding system of equations are available, and the numerical solution
should be significantly more difficult than the one presented above. Note
that inhomogeneous broadening was mentioned in previous publications on the
phonon bottleneck (see, e.g., Ref. \cite{abrble70}) but it was treated as
part of the ``phenomenological'' line width of the transition, without any
frequency resolution of the type suggested in this paragraph.

Spin-spin interactions make the problem even more complicated. Although the
effects of inhomogeneous broadening and spin-spin interaction are important
for comparison with experiment, it does not make sense to consider them in
the present paper for the reason stressed above. In principle, to be on a
safe ground, one has to find a satisfactory solution for the core problem of
the PB and obtain a correct line shape of the emitted phonons, before adding
inhomogeneous broadening and other real-life effects to the model.

Of course, one can argue that in the presence of inhomogeneous broadening
the phonon line shape should be dominated by the latter and one obtains a
good-looking everywhere positive line shape by a kind of averaging over the
spin transition frequency $\omega _{i}$, instead of the unsatisfactory line
shape in Fig.\ \ref{Fig-nk}b. According to this logic, including
inhomogeneous broadening would have more sense, as the next step, than
struggling for the accurate line shape in the bare model. Practically this
might be true, although one cannot easily find a microscopic justification
for such an approach.

It should be stressed that the whole consideration in this paper pertains to
magnetically diluted systems, in which the phases of phonons emitted by one
spin and reaching another spin are random. This excludes magnetically dense
systems such as molecular magnets. In the latter, emission of phonons by
different spins is correlated, that in the ideal case leads to the phonon
superradiance.\cite{chugar04prl} The interplay of the phonon bottleneck and
phonon superradiance is an exciting and challenging problem, and its
theoretical description requires including further correlators that have
been ignored above. Obviously, for magnetically dense systems applicability
of the \emph{ad} \emph{hoc} bottleneck equations\cite
{faustr61jpcs,scojef62pr,brywag67pr,abrble70,pinfai90prb} is even less
justified than for the diluted systems.

\vspace{-0.5cm}
\begin{acknowledgements}The author gratefully acknowledges stimulating discussions with
E. M. Chudnovsky and A. Kent. \end{acknowledgements}


\end{document}